# Subjective and Objective Quality Assessment of Image: A Survey

Pedram Mohammadi[*], Abbas Ebrahimi-Moghadam[*,1], and Shahram Shirani[**]

[*]*Department of Electrical Engineering*
*Ferdowsi University of Mashhad, Mashhad, Iran*
[**]*Department of Electrical and Computer Engineering*
*McMaster University, Hamilton, ON, Canada*

**Abstract**

With the increasing demand for image-based applications, the efficient and reliable evaluation of image quality has increased in importance. Measuring the image quality is of fundamental importance for numerous image processing applications, where the goal of image quality assessment (IQA) methods is to automatically evaluate the quality of images in agreement with human quality judgments. Numerous IQA methods have been proposed over the past years to fulfill this goal. In this paper, a survey of the quality assessment methods for conventional image signals, as well as the newly emerged ones, which includes the high dynamic range (HDR) and 3-D images, is presented. A comprehensive explanation of the subjective and objective IQA and their classification is provided. Six widely used subjective quality datasets, and performance measures are reviewed. Emphasis is given to the full-reference image quality assessment (FR-IQA) methods, and 9 often-used quality measures (including mean squared error (MSE), structural similarity index (SSIM), multi-scale structural similarity index (MS-SSIM), visual information fidelity (VIF), most apparent distortion (MAD), feature similarity measure (FSIM), feature similarity measure for color images ($FSIM_C$), dynamic range independent measure (DRIM), and tone-mapped images quality index (TMQI)) are carefully described, and their performance and computation time on four subjective quality datasets are evaluated. Furthermore, a brief introduction to 3-D IQA is provided and the issues related to this area of research are reviewed.

*Key words*: Image quality assessment (IQA), high dynamic range (HDR) images, 3-D image quality assessment, full-reference IQA, reduced-reference IQA, no-reference IQA

---

[1]Corresponding author
*Email addresses:* pedram.mohammadi@stu.um.ac.ir (Pedram Mohammadi), a.ebrahimi@um.ac.ir (Abbas Ebrahimi-Moghadam), shirani@mcmaster.ca (Shahram Shirani)

# 1 Introduction

Digital images are rapidly finding their way into our daily lives due to the explosion of information in the form of visual signals. These images often pass through several processing stages before they reach to their end-users. In most cases, these end-users are human observers. Through different processing stages, e.g., acquisition, compression, and transmission, images are subjected to different types of distortions which degrade the quality of them. For example, in image compression, lossy compression schemes introduce blurring and ringing effects, which leads to quality degradation. Moreover, in the transmission stage, due to limited bandwidth of the channels, some data might be dropped, which results in quality degradation of the received image.

In order to maintain, control, and enhance the quality of images, it is essential for image communication, management, acquisition, and processing systems to assess the quality of images at each stage. IQA plays an important role in visual signal communication and processing. The application scope of IQA includes, but is not confined to, image acquisition [1], segmentation [2], printing and display systems [3,4], image fusion [5], and biomedical imaging [6,7]. IQA methods can be categorized into subjective and objective methods.

Since human observers are the ultimate users in most of the multimedia applications, the most accurate and also reliable way of assessing the quality of images is through subjective evaluation. However, subjective evaluations are expensive and time consuming, which makes them impractical in real-world applications. Moreover, subjective experiments are further complicated by many factors including viewing distance, display device, lighting condition, subjects' vision ability, and subjects' mood. Therefore, it is necessary to design mathematical models that are capable of predicting the quality evaluation of an average human observer.

The goal of objective IQA is to design mathematical models that are able to predict the quality of an image accurately and automatically. An ideal objective IQA method should be able to mimic the quality predictions of an average human observer. Based on the availability of a reference image which is considered to be distortion-free and have perfect quality, the objective quality assessment methods can be classified into three categories. The first category is full-reference image quality assessment (FR-IQA) where the undistorted, perfect quality reference image is fully available. The second category is reduced-reference image quality assessment (RR-IQA) where the reference image is not fully available. Instead, some features of the reference image are extracted and employed as side information in order to evaluate the quality of the test image. The third category is no-reference image quality assessment (NR-IQA) in which we don't have access to the reference image. Since in many real-world applications the reference image is not accessible, NR-IQA methods are very convenient in practice.

This paper intends an overview of the subjective and objective IQA. Classification of both subjective and objective IQA is presented. Six widely used subjective quality datasets and performance measures are reviewed. Emphasis is given to FR-IQA measures and 9 often-used quality measures (including MSE, SSIM [8], MS-SSIM [9], VIF [10], MAD [11], FSIM [12], FSIM$_C$ [12], DRIM [13], and TMQI [14]) are carefully described, and the computation time and performance of these methods are evaluated on four subjective datasets. Moreover, a brief introduction to 3-D IQA is provided, issues associated with this field are presented, some of the objective 3-D IQA methods are briefly reviewed, and some 3-D image datasets are introduced. This paper is organized as follows:

In section 2, subjective IQA is reviewed, recommendations on designing subjective experiments are presented, and four common standardized subjective IQA measures are reviewed. In section 3, objective IQA and its three main categories are reviewed. Moreover, a detailed description of six FR-IQA methods for gray-scale images (including MSE, SSIM, MS-SSIM, VIF, MAD, and FSIM) is provided. In section 4, a brief introduction to color images quality assessment is presented, and one FR-IQA method for color images (namely FSIM$_C$) is described. In section 5, a brief introduction to HDR images quality assessment is provided. Moreover, two FR-IQA methods for images with different dynamic ranges (including DRIM and TMQI) are explained in details. In section 6, six widely used subjective quality datasets and performance measures are summarized. In section 7, performance and computation time of FR-IQA measures described in previous sections are evaluated on four subjective quality datasets (including LIVE dataset [15], CSIQ dataset [16], TID2008 dataset [17], and the dataset presented in [18]). In section 8, a concise introduction to 3-D IQA is provided, some of the objective 3-D IQA methods are briefly reviewed, and some 3-D image datasets are presented. Finally, section 9 concludes the paper.

## 2 Subjective image quality assessment

The most reliable method for assessing the quality of images is through subjective testing, since human observers are the ultimate users in most of the multimedia applications. In subjective testing a group of people are asked to give their opinion about the quality of each image. In order to perform a subjective image quality testing, several international standards are proposed [19-25] which provide reliable results. Here, we briefly describe some of these international standards:

ITU-R BT.500-11 [19] proposes different methods for subjective quality assessment of television pictures. This is a widely used standard, which contains information about viewing condition, instructions on how to perform subjective experiments, test materials, and presentation of subjective results.

ITU-T P.910 [21] proposes the standard method for digital video quality assessment with transmission rate below 1.5 Mbits/sec.

ITU-R BT.814-1 [22] is proposed in order to set the brightness and contrast of the display devices.

ITU-R BT.1129-2 [23] is proposed for assessing the quality of the standard definition (SD) video sequences.

In the following subsections, we will briefly describe some of the standardized subjective IQA methods.

## 2.1. Single stimulus categorical rating

In this method, test images are displayed on a screen for a fixed amount of time, after that, they will disappear from the screen and observers will be asked to rate the quality of them on an abstract scale containing one of the five categories: excellent, good, fair, poor, or bad. All of the test images are displayed randomly. In order to avoid quantization artifacts, some methods use continuous rather than categorical scales [19].

## 2.2. Double stimulus categorical rating

This method is similar to single stimulus method. However, in this method both the test and reference images are being displayed for a fixed amount of time. After that, images will disappear from the screen and observers will be asked to rate the quality of the test image according to the abstract scale described earlier.

## 2.3. Ordering by force-choice pair-wise comparison

In this type of subjective assessment, two images of the same scene are being displayed for observers. Afterward, they are asked to choose the image with higher quality. Observers are always required to choose one image even if both images possess no difference. There is no time limit for observers to make the decision. The drawback of this approach is that it requires more trials to compare each pair of conditions [26]. In [27,28], two methods for reducing the number of trials are proposed.

## 2.4. Pair-wise similarity judgments

As we mentioned before, in force-choice comparison, observers are required to choose one image even if they see no difference between the pair of images. However, in pair-wise similarity judgment observers are asked not only to choose the image with higher quality, but also to indicate the level of difference between them on a continuous scale.

One might be tempted to use the raw rating results such as: excellent, good, fair, and etc. for quality scores. However, these rating results are unreliable. One reason for this is that observers are likely to assign different quality scales to each scene and even distortion types [29]. Here, we briefly introduce two scoring methods used in the subjective IQA.

## 2.5. Difference mean opinion score (DMOS)

Instead of directly applying rating results, modern IQA metrics use differences in quality between images. DMOS is defined as the difference between the raw quality score of the reference and test images. DMOS is calculated using the following equation:

$$d_{i,j} = r_{i,ref}(j) - r_{i,j} \qquad (1)$$

where $r_{i,j}$ is the raw score for the $i-th$ subject and the $j-th$ image. Also, $r_{i,ref}(j)$ denotes the raw score given by the $i-th$ subject to the reference image corresponding to the $j-th$ test image.

2.6. *Z-score*

In order to easily compare each observer's opinion about the quality of images, a linear transform that makes the mean and variance equal for all observers is employed. The outcome of such transform is called Z-score and it can be computed using the following equation:

$$z_{i,j} = \frac{d_{i,j} - \bar{d}_i}{\sigma_i} \qquad (2)$$

The mean DMOS, $\bar{d}_i$, and the standard deviation, $\sigma_i$, are computed across all images that are rated by the $i-th$ subject.

Subjective quality assessment methods provide accurate and reliable measurements of the quality of visual signals. However, these methods suffer from different drawbacks that limits their applications:

- They are time consuming and expensive. This is due to the fact that subjective results are obtained through experiments with many observers.
- They cannot be incorporated into real-time applications such as image compression, and transmission systems.
- Their results depend heavily on the subjects' physical conditions and emotional state. Moreover, other factors such as display device and lighting condition affect the results of such experiments.

Therefore, it is necessary to design mathematical models that are able to predict the perceptual quality of visual signals in a consistent manner with subjective evaluations.

## 3 Objective image quality assessment

The goal of objective IQA is to design mathematical models that are able to predict the quality of an image accurately and also automatically. An ideal objective IQA method should be able to mimic the quality predictions of an average human observer. Objective IQA methods have a wide variety of applications [30]:

- They can be used to monitor image quality in quality control systems. For example, image acquisition systems can employ an objective IQA metric to monitor and automatically adjust themselves in order to obtain the best quality image data.

- They can be used to benchmark image processing algorithms. For example, if a number of image enhancement algorithms are available, an objective IQA metric can be employed to choose the algorithm that provides the higher quality images.
- They can be used to optimize image processing and transmission systems. For example, in a visual communication network, an objective IQA metric can be employed to optimize pre-filtering and bit assignment algorithms at the encoder and post-filtering and reconstruction algorithms at the decoder.

Based on the availability of a distortion-free, perfect quality reference image, objective IQA methods can be classified into three categories. The first category is full-reference image quality assessment (FR-IQA) where the reference image is fully available. The second category is reduced-reference image quality assessment (RR-IQA) where only partial information about the reference image is available. And the third category is no-reference image quality assessment (NR-IQA) where neither the reference image nor its features are available for quality evaluation.

Objective IQA methods can also be categorized based on their application scope [30]. General purpose methods are the ones that do not assume a specific distortion type. Therefore, these methods are useful in a wide range of applications. On the other hand, application specific methods are the ones that are designed for specific distortion types. An example of these methods are the algorithms designed for image compression applications. Many quality metrics in image compression are designed for block-DCT or wavelet-based image compression.

In the following subsections, the characteristics of the three main categories of the objective IQA are described.

3.1. *No-reference image quality assessment (NR-IQA)*

In many real-world applications, such as image communication systems, the reference image is not available and the quality evaluation is solely based on the test image. NR-IQA is a more difficult task in comparison to RR-IQA and FR-IQA methods. However, human beings usually can efficiently assess the quality of a test image without using any reference image. This is probably due to the fact that our brain holds a lot of information about how an image should or should not look like in real world [30]. Some NR-IQA methods can be found in [31-36].

3.2. *Reduced-reference image quality assessment (RR-IQA)*

In RR-IQA, the reference image is not fully accessible. Instead, a number of features are extracted from the reference image. These features are employed by the quality assessment method as the side information for evaluating the quality of the test image. RR-IQA methods can be employed in a number of applications. For instance, they can be used to track the level of visual quality degradation of image and video data transmitted via real-time visual communication networks.

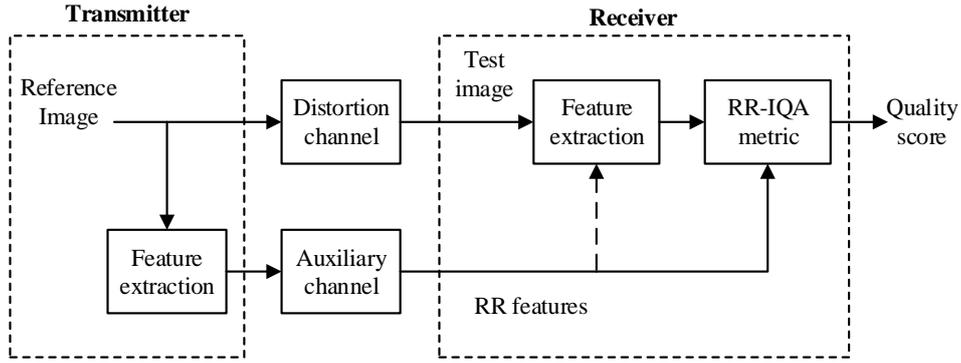

**Fig. 1.** The framework of an RR-IQA system.

Fig.1 shows the framework of an RR-IQA system. At the transmitter, a feature extraction process extracts certain features from the reference image and transmits them through an auxiliary channel. Feature extraction process is also applied to the test image at the receiver. The feature extraction process at the receiver can also be adopted to the side information at the receiver, which is shown as a dashed arrow in the figure, or it can be the same as in the transmitter. In order to obtain a single score for the overall quality of the test image, the features extracted from both, the reference and test images, are employed. An important parameter in the design of an RR-IQA system is the data rate used to encode the side information. If a high data rate is available, it is possible to include more information about the reference image, which allows more accurate quality predictions. If the data rate is high enough that all the information about the reference image can be transmitted, then the RR-IQA metric can be considered as a FR-IQA metric. On the other hand, if a low data rate is used, then only a small amount of information about the reference image can be transmitted. This results in less accurate quality predictions. In the case of zero data rate, the RR-IQA metric is considered as an NR-IQA metric. In real-world RR-IQA systems, the maximally allowed data rate is usually low [30]. Limited values for the data rate limits the feature selection process in RR-IQA systems. Therefore, selected features should satisfy following criteria:

- They should be able to provide an efficient summary of the reference image.
- They should be sensitive to a variety of distortion types.
- They should possess good perceptual relevance.

On the basis of design philosophy, RR-IQA methods can be loosely classified in three categories [37]:

3.2.1. *Methods based on the models of the image source*

The methods of this type are often statistical models that capture a priori of low level statistical features of natural images. These methods often have a low data rate. This is due to the fact that the parameters of these methods are able to summarize the image information in an efficient manner. Some of the methods in this category can be found in [37-40].

3.2.2. *Methods based on capturing image distortions*

The methods in this category are most useful when sufficient information about the image distortions is available. The application scope of these methods is limited, since they are unable to capture the distortions that they are not designed for. Some of the methods in this category can be found in [41-44].

3.2.3. *Methods based on the models of human visual system*

In designing the methods in this category, physiological and/or psychophysical studies may be employed. These methods have shown good performance for JPEG and JPEG2000 compression schemes. Some of the methods in this category can be found in [45,46].

3.3. *Full-reference image quality assessment (FR-IQA)*

In FR-IQA metrics, the perfect quality reference image is fully available for quality prediction process. The application scope of these metrics includes image compression [47], watermarking [48,49], and so on. In the following subsections, we will comprehensively describe six FR-IQA methods. The selected methods are widely cited in the literature, and have been reported to have good performance by researchers. Moreover, the authors of the selected metrics have released the source codes of their respective metrics. Therefore, results of the selected metrics are easy to reproduce. The six FR-IQA metrics described in the following subsections include mean squared error (MSE), structural similarity index (SSIM) [8], multi-scale structural similarity index (MS-SSIM) [9], visual information fidelity (VIF) [10], most apparent distortion (MAD) [11], and feature similarity index (FSIM) [12]. It is important to note that all of these six quality evaluation metrics are designed for gray-scale images.

In all of the following subsections, $I_{ref}$ and $I_{tst}$ denote the reference and test images respectively, and subscript *ref* denotes reference and *tst* test. Moreover, $W$ and $H$ represent the width and height of images respectively.

3.3.1. *Mean squared error (MSE)*

MSE denotes the power of the distortion, i.e., the difference between the reference and test images. MSE value can be calculated using the following equation:

$$\text{MSE} = \frac{1}{WH} \sum_{j=1}^{H} \sum_{i=1}^{W} \left( I_{ref}(i,j) - I_{tst}(i,j) \right)^2 \tag{3}$$

MSE is often converted to peak-signal-to-noise ratio (PSNR). PSNR is the ratio of maximum possible power of a signal and power of distortion, and it is calculated by:

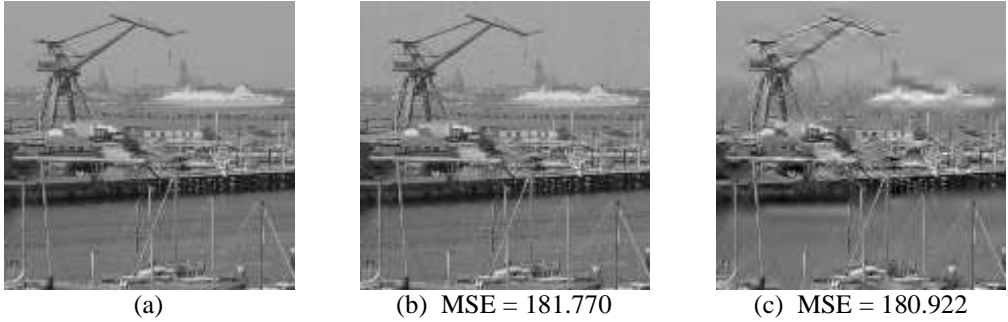

(a)          (b) MSE = 181.770          (c) MSE = 180.922

**Fig. 2.** *Harbor* image altered with two types of distortions: (a) reference image; (b) white Gaussian noise; (c) quantization of the LH subbands of a 5-level DWT of the image with equal distortion contrast at each scale. All images are extracted from [51].

$$\text{PSNR} = 10\log\left(\frac{D^2}{\text{MSE}}\right) \qquad (4)$$

where $D$ denotes the dynamic range of pixel intensities, e.g., for an 8 bits/pixel image we have $D = 255$.

MSE possesses some characteristics that make it a widely used performance measure in the field of signal processing. Following are some of these characteristics [50]:

- It is a simple, computationally inexpensive method.
- It has a physically clear meaning, i.e., it is a natural way of defining the energy of an error signal.
- Since MSE satisfies properties like convexity, symmetry, and differentiability, it is considered as an excellent measure in optimization applications.
- It is considered as a convention, i.e., it is extensively used for optimization and assessment in a wide range of signal processing applications.

Despite the above interesting features of MSE, when it comes to predicting human perception of image quality, MSE shows poor performance. This is due to the fact that some of the important physiological and psychophysical characteristics of the human visual system (HVS) are not accounted for by this measure.

An instructive example is shown in Fig. 2, where the reference image (a) is altered by two types of distortions: white Gaussian noise (b), and quantization of the LH subbands of a 5-level discrete wavelet transform (DWT) of the image with equal distortion contrast at each scale (c). It is important to note that images (b) and (c) have nearly similar MSE values. However, they have different visual qualities. There exist some implicit assumptions when using the MSE measure which makes it a poor measure of image quality. These assumptions are listed as follows [50]:

- If the reference and test images are randomly re-ordered in a similar manner, the MSE between them will remain unchanged. This demonstrates that MSE is independent of temporal or spatial relationship between samples of the reference image.

- For a specific distortion signal, MSE remains unchanged regardless of which reference signal it is added to.
- MSE is independent of the error signal samples' sign.
- Image signals are considered equally important when MSE is computed.

### 3.3.2. *Structural similarity index (SSIM)*

The SSIM algorithm [8] assumes that HVS is highly adapted for extracting structural information from a scene. Therefore, this algorithm attempts to model the structural information of an image. The SSIM algorithm is based on the fact that pixels of a natural image demonstrate strong dependencies and these dependencies carry useful information about the structure of a scene. Therefore, a method that is capable of measuring structural information change can provide a good approximation of perceived image distortion. The SSIM algorithm defines image degradation as perceived change in structural information. In [8], it is stated that the structure of the objects in a scene is independent of local luminance and contrast. Therefore, to extract the structural information, we should separate the effect of illumination. In this algorithm, structural information in an image is defined as those traits that represent the structure of objects in that image, independent of the local luminance and contrast.

The SSIM algorithm performs similarity measurement in three steps: luminance comparison, contrast comparison, and structure comparison:

First, the luminance of each image signal is compared. The estimated mean intensity is computed as follows:

$$\mu_{ref} = \frac{1}{WH} \sum_{j=1}^{H} \sum_{i=1}^{W} I_{ref}(i,j). \tag{5}$$

The luminance comparison function, $l(I_{ref}, I_{tst})$, is a function of $\mu_{ref}$ and $\mu_{tst}$. Second, the contrast of each image signal is compared. For estimating the contrast, standard deviation is being used. An unbiased estimate of standard deviation in discrete form is as follows:

$$\sigma_{ref} = \left( \frac{1}{WH-1} \sum_{j=1}^{H} \sum_{i=1}^{W} \left( I_{ref}(i,j) - \mu_{ref} \right)^2 \right)^{\frac{1}{2}} \tag{6}$$

The contrast comparison function, $c(I_{ref}, I_{tst})$, is a function of $\sigma_{ref}$ and $\sigma_{tst}$. Third, the structure of each image signal is compared. Structure comparison function, $s(I_{ref}, I_{tst})$, is a function of $[I_{ref} - \mu_{ref}]/\sigma_{ref}$ and $[I_{tst} - \mu_{tst}]/\sigma_{tst}$. Finally, three comparison functions are combined and an overall similarity measure is produced. The overall similarity measure, $S(I_{ref}, I_{tst})$, is a function of $l(I_{ref}, I_{tst})$, $c(I_{ref}, I_{tst})$, and $s(I_{ref}, I_{tst})$. The function $S(I_{ref}, I_{tst})$ satisfies following conditions:

- Symmetry: $S(\mathbf{I}_{ref}, \mathbf{I}_{tst}) = S(\mathbf{I}_{tst}, \mathbf{I}_{ref})$.
- Boundedness: $-1 \leq S(\mathbf{I}_{ref}, \mathbf{I}_{tst}) \leq 1$.
- Unique maximum: $S(\mathbf{I}_{ref}, \mathbf{I}_{tst}) = 1$, if and only if $\mathbf{I}_{ref} = \mathbf{I}_{tst}$.

Definitions of $l(\mathbf{I}_{ref}, \mathbf{I}_{tst})$, $c(\mathbf{I}_{ref}, \mathbf{I}_{tst})$, and $s(\mathbf{I}_{ref}, \mathbf{I}_{tst})$, are as follows:

For luminance comparison function we have:

$$l(\mathbf{I}_{ref}, \mathbf{I}_{tst}) = \frac{2\mu_{ref}\mu_{tst} + T_1}{\mu_{ref}^2 + \mu_{tst}^2 + T_1} \quad (7)$$

where $T_1$ is a positive stabilizing constant chosen to prevent the denominator from becoming too small. We have:

$$T_1 = (t_1 D)^2 \quad (8)$$

where $D$ is the dynamic range of pixel values and $t_1 \ll 1$ is a small constant. For contrast comparison function we have:

$$c(\mathbf{I}_{ref}, \mathbf{I}_{tst}) = \frac{2\sigma_{ref}\sigma_{tst} + T_2}{\sigma_{ref}^2 + \sigma_{tst}^2 + T_2} \quad (9)$$

where $T_2 = (t_2 D)^2$ is a positive stabilizing constant. And $t_2 \ll 1$. For structure comparison function we have:

$$s(\mathbf{I}_{ref}, \mathbf{I}_{tst}) = \frac{\sigma_{ref,tst} + T_3}{\sigma_{ref}\sigma_{tst} + T_3} \quad (10)$$

where $T_3$ is a positive stabilizing constant. In (10), $\sigma_{ref,tst}$ is the correlation coefficient between the reference and test images. In the discrete form, $\sigma_{ref,tst}$ can be estimated by:

$$\sigma_{ref,tst} = \frac{1}{WH-1} \sum_{j=1}^{H} \sum_{i=1}^{W} \left(\mathbf{I}_{ref}(i,j) - \mu_{ref}\right)\left(\mathbf{I}_{tst}(i,j) - \mu_{tst}\right) \quad (11)$$

Finally, structural similarity index is defined as:

$$\text{SSIM}(\mathbf{I}_{ref}, \mathbf{I}_{tst}) = \left[l(\mathbf{I}_{ref}, \mathbf{I}_{tst})\right]^\alpha \left[c(\mathbf{I}_{ref}, \mathbf{I}_{tst})\right]^\beta \left[s(\mathbf{I}_{ref}, \mathbf{I}_{tst})\right]^\gamma \quad (12)$$

where $\alpha$, $\beta$, and $\gamma$ are positive constants chosen to indicate the relative importance of each component. The universal quality index (UQI) [52,53] is a special case of the SSIM index when: $T_1 = T_2 = T_3 = 0$ and $\alpha = \beta = \gamma = 1$. Since image statistical features and distortions are usually space-variant, authors of [8] employ the SSIM index locally instead of globally. Another reason for this is that by applying the SSIM index locally, a quality map of the image which conveys more information about the quality degradation can be generated.

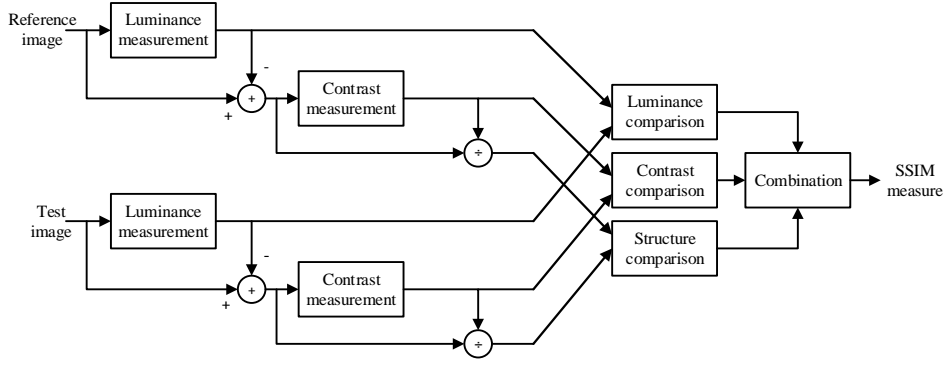

**Fig. 3.** The block diagram of the SSIM algorithm.

In order to achieve this, authors used an $11\times11$ circular symmetric Gaussian weighting function $w = \{w_{i,j} \mid i = 1, 2, ..., W \text{ and } j = 1, 2, ..., H\}$ with standard deviation of 1.5 samples, normalized to unit sum $\sum_{j=1}^{H}\sum_{i=1}^{W} w_{i,j} = 1$. Using this function, the estimates of local statistics $\mu_{ref}$, $\sigma_{ref}$, and $\sigma_{ref,tst}$ are calculated as follows:

$$\mu_{ref} = \sum_{j=1}^{H}\sum_{i=1}^{W} w_{i,j} I_{ref}(i,j) \tag{13}$$

$$\sigma_{ref} = \left(\sum_{j=1}^{H}\sum_{i=1}^{W} w_{i,j}\left(I_{ref}(i,j) - \mu_{ref}\right)^2\right)^{\frac{1}{2}} \tag{14}$$

$$\sigma_{ref,tst} = \sum_{j=1}^{H}\sum_{i=1}^{W} w_{i,j}\left(I_{ref}(i,j) - \mu_{ref}\right)\left(I_{tst}(i,j) - \mu_{tst}\right) \tag{15}$$

In order to have a single overall quality measure for the entire image, authors of [8] use a mean SSIM (MSSIM) index to evaluate the overall quality:

$$\text{MSSIM}(I_{ref}, I_{tst}) = \frac{1}{M_w}\sum_{i=1}^{M_w} SSIM(I_{ref}^{i}, I_{tst}^{i}) \tag{16}$$

where $M_w$ is the total number of local windows, and $I_{ref}^{i}$ and $I_{tst}^{i}$ are image contents at the $i-th$ local window. The block diagram of the SSIM algorithm is presented in Fig. 3. Some applications of the SSIM algorithm are image fusion [5], image watermarking [54], remote sensing [55], and visual surveillance [56].

3.3.2.1. *Parameter specification in the SSIM algorithm*

There are several parameters in the SSIM algorithm that need to be specified. First, for computing (8) the values of $t_1$ and $D$ are set to be 0.01 and 255 respectively. Second, for computing (9) the value of $t_2$ is set to be 0.03. Third, in

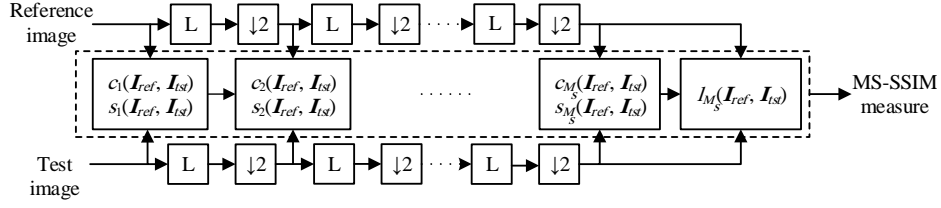

**Fig. 4.** The block diagram of the MS-SSIM algorithm. L: low-pass filter; ↓2: downsampling by factor of 2.

(10) we have: $T_3 = T_2/2$. It is stated in [8] that the performance of the SSIM algorithm is fairly insensitive to the values of $T_1$, $T_2$, and $T_3$. Finally, in order to simplify (12), SSIM algorithm sets $\alpha = \beta = \gamma = 1$.

Authors of [8] have provided a MATLAB implementation of the SSIM algorithm that is available at [57].

### 3.3.3. *Multi-scale structural similarity index (MS-SSIM)*

The SSIM algorithm described earlier is considered a single-scale approach that achieves its best performance when applied at an appropriate scale. Moreover, choosing the right scale depends on the viewing conditions, e.g., viewing distance and the resolution of the display. Therefore, this algorithm lacks the ability to adapt to these conditions. This drawback of the SSIM algorithm motivated researchers to design a multi-scale structural similarity index (MS-SSIM) [9]. The advantage of the multi-scale methods, like MS-SSIM, over single-scale methods, like SSIM, is that in multi-scale methods image details at different resolutions and viewing conditions are incorporated into the quality assessment algorithm. The block diagram of the MS-SSIM algorithm is presented in Fig. 4. After taking the reference and test images as input, this algorithm performs low-pass filtering and downsampling (by factor of 2) in an iterative manner. At each scale, (9) and (10) are calculated. However, (7) is computed only at $M_s-th$ scale. The final MS-SSIM index is calculated using the following equation:

$$\text{MS-SSIM}(\mathbf{I}_{ref}, \mathbf{I}_{tst}) = \left[l_{M_s}(\mathbf{I}_{ref}, \mathbf{I}_{tst})\right]^{\alpha_{M_s}} \cdot \prod_{i=1}^{M_s} \left[c_i(\mathbf{I}_{ref}, \mathbf{I}_{tst})\right]^{\beta_i} \left[s_i(\mathbf{I}_{ref}, \mathbf{I}_{tst})\right]^{\gamma_i} \quad (17)$$

where $c_i(\mathbf{I}_{ref}, \mathbf{I}_{tst})$ and $s_i(\mathbf{I}_{ref}, \mathbf{I}_{tst})$ are the contrast and the structure comparison function at the $i-th$ scale respectively, and $l_{M_s}(\mathbf{I}_{ref}, \mathbf{I}_{tst})$ is the luminance comparison function at the $M_s-th$ scale. Moreover, $\alpha_{M_s}$, $\beta_i$, and $\gamma_i$ are positive constants chosen to indicate the relative importance of each component. In [9], $\alpha_i = \beta_i = \gamma_i$ for all $j$, and $\sum_{i=1}^{M_s} \gamma_i = 1$.

3.3.3.1. *Parameter specification in the MS-SSIM algorithm*

An image synthesis-based approach is used in order to calculate the exponents of (17). In [9], for a given original 8 bits/pixel gray-scale test image, a matrix of test images is constructed. Each element in the matrix is an image that is related to a specific MSE value and a specific scale. Each test image in the matrix is created by randomly adding white noise to the original test image. 5 scales and 12 distortion levels are used that yields a matrix of total of 60 images. Moreover, 10 original test images of size $64 \times 64$ with different contents are used in order to create 10 sets of test images, resulting in the total number of test images to be 600. As mentioned in [9], 8 subjects (including one of the authors) have participated in the subjective experiment for calculating the exponents of (17). Subjects had general understanding of the human vision, but were unaware of the goal of the experiment. After seeing all 10 sets of test images in a fixed viewing distance, they were asked to choose one image in each of 5 scales that they think have the same quality. After that, the positions of chosen images in each scale is saved and averaged across all test images and subjects. Test results are then normalized so that their sum becomes equal to 1. The resulting exponents for each of 5 scales are: $\beta_1 = \gamma_1 = 0.0448$, $\beta_2 = \gamma_2 = 0.2856$, $\beta_3 = \gamma_3 = 0.3001$, $\beta_4 = \gamma_4 = 0.2363$, $\beta_5 = \gamma_5 = \alpha_5 = 0.1333$.

Authors of [9] have provided a MATLAB implementation of the MS-SSIM algorithm that is available at [58].

3.3.4. *Visual information fidelity (VIF)*

VIF algorithm [10] models natural images in the wavelet domain using Gaussian scale mixtures (GSMs). Images and videos that are taken from natural environment by using high quality capturing devices operating in visual spectrum are classified as natural scenes. For a review of natural scene models see [59]. VIF algorithm consists of three components: source model, distortion model, and HVS model.

3.3.4.1. *Source model*

As stated earlier, VIF algorithm models natural images in wavelet domain using GSM model. A GSM is defined as a random field that can be determined as a product of two independent random fields [60]. In other words, a GSM like *c* can be expressed as:

$$\boldsymbol{c} = z\boldsymbol{u} \qquad (18)$$

where $z$ is a random field containing positive scalars, and $\boldsymbol{u}$ is a Gaussian vector random field with zero mean and covariance $\mathbf{C}_u$. In [10], it is assumed that $\boldsymbol{u}$ consists of independent components. VIF algorithm models each subband of image's wavelet decomposition as a GSM random field. Each subband coefficients are grouped into non-overlapping blocks of size $M_n$.

3.3.4.2. *Distortion model*

Distortion is modeled in the wavelet domain as signal attenuation and additive noise. This model is defined as follows:

$$d = gc + v \tag{19}$$

where $c$ is a random field from a subband in the reference image, $d$ is a random field from a subband in the test image, $g$ is a deterministic scalar field, and $v$ is a random field from a stationary white additive Gaussian noise with zero mean and covariance $\mathbf{C}_v = \sigma_v^2 \mathbf{I}$. In [10], random fields $v$, $z$, and $u$ are assumed to be independent from one another. Moreover, random field $g$ is considered to be slow varying.

3.3.4.3. *HVS model*

HVS is modeled as a distortion channel that adds noise to the input signal, limiting the amount of information that flows through the channel. This visual noise is characterized as a zero mean stationary additive white Gaussian noise modeled in the wavelet domain. HVS noise is modeled as stationary random fields $n'$ and $n$ which are zero mean, uncorrelated multivariate Gaussians with the same covariance ($\mathbf{C}_n = \mathbf{C}_{n'} = \sigma_n^2 \mathbf{I}$, where $\sigma_n^2$ is considered the variance of the visual noise). The outputs of the HVS channels are as follows:

$$e = c + n \tag{20}$$

$$f = d + n' \tag{21}$$

where $e$ is the output of the HVS channel when the input is the reference image and $f$ is the output of the same channel when the input is the test image. Random fields $n$ and $n'$ are assumed to be independent of $u$, $z$, and $v$.

With the source, distortion, and HVS models described earlier, the VIF quality measure can be calculated. Consider $C = \{c_1, c_2, ..., c_{M_r}\}$ and $z = \{z_1, z_2, ..., z_{M_r}\}$ to be a collection of $M_r$ realization from the random field $c$ and $z$ respectively. Moreover, let $D$, $E$, and $F$ be defined in a similar manner in terms of $d$, $e$, and $f$. In [10], all GSM vectors are constructed from a non-overlapping $3 \times 3$ neighborhoods. In order to calculate the VIF measure, information content of the reference and the test images needs to be calculated.

3.3.4.4. *Calculating reference image's information*

The amount of information that can be extracted from a particular subband in the reference image, $I(C; E | z)$, is calculated as follows:

$$I(C;E|z) = \sum_{i=1}^{M_r}\left[h(c_i + n_i|z_i) - h(n_i|z_i)\right]$$
$$= \frac{1}{2}\sum_{i=1}^{M_r}\log_2\left(\frac{|z_i^2 C_u + \sigma_n^2 I|}{|\sigma_n^2 I|}\right) \quad (22)$$

where $h(\cdot)$ and $|\cdot|$ denote the differential entropy of a continuous random vector and determinant operator respectively. Since $C_u$ is symmetric, by using matrix factorization we can write it as $C_u = Q\Lambda Q^T$, where $Q$ is an orthonormal matrix, and $\Lambda$ is a diagonal matrix containing eigenvalues $\lambda_k$. Using this factorization, (22) can be written as:

$$I(C;E|z) = \frac{1}{2}\sum_{i=1}^{M_r}\sum_{k=1}^{M_e}\log_2\left(1 + \frac{z_i^2 \lambda_k}{\sigma_n^2}\right) \quad (23)$$

where $M_e$ is the total number of eigenvalues in $\Lambda$.

3.3.4.5. *Calculating test image's information*

The amount of information that can be extracted from a particular subband in the test image, $I(C;F|z)$, is calculated as follows:

$$I(C;F|z) = \sum_{i=1}^{M_r}\left[h(g_i c_i + v_i + n_i|z_i) - h(v_i + n_i|z_i)\right]$$
$$= \frac{1}{2}\sum_{i=1}^{M_r}\log_2\left(\frac{|g_i^2 z_i^2 C_u + (\sigma_{v,i}^2 + \sigma_n^2)I|}{|(\sigma_{v,i}^2 + \sigma_n^2)I|}\right). \quad (24)$$

Using the same factorization as before, (24) can be written as:

$$I(C;F|z) = \frac{1}{2}\sum_{i=1}^{M_r}\sum_{k=1}^{M_e}\log_2\left(1 + \frac{g_i^2 z_i^2 \lambda_k}{\sigma_{v,i}^2 + \sigma_n^2}\right) \quad (25)$$

It has been discovered in [10] that the ratio of equations (23) and (25) relates well with visual quality. Therefore, by using the assumption that each subband is completely independent of others in terms of their respective random fields as well as the distortion model parameters, the VIF quality measure is calculated as follows:

$$\text{VIF} = \frac{\sum_{j\in subbands} I(C^j;F^j|z^j)}{\sum_{j\in subbands} I(C^j;E^j|z^j)} \quad (26)$$

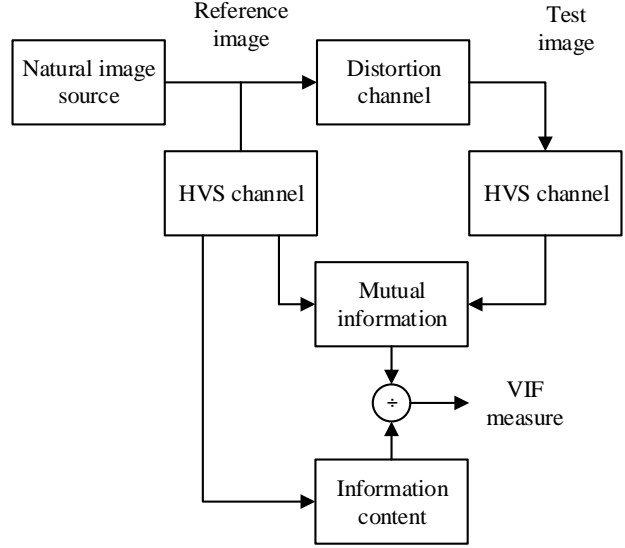

**Fig. 5.** The block diagram of the VIF algorithm.

where $j$ is the subband index, and $I\left(C^{j}; F^{j} \mid z^{j}\right)$ and $I\left(C^{j}; E^{j} \mid z^{j}\right)$ are the corresponding mutual information of the $j-th$ subband. In [10], summation is performed over subbands at the finest scale. The VIF measure can be calculated by using an entire subband of image or by using a spatially localized region of subband coefficients. In the first case, VIF measure is a single number that quantifies the overall quality of the image, and in the second case, a sliding window could be used to obtain a quality map of the image. The block diagram of the VIF algorithm is presented in Fig. 5.

For all practical distortion types, the VIF measure takes its values in the interval $[0,1]$. $VIF = 0$ means that all the information about the reference image has been lost due to presence of distortions. For images with higher perceptual quality, the value of the VIF measure is close to 1. A linear contrast enhancement of the reference image, that doesn't add distortion to it, results in the VIF measure greater than 1. Therefore, $VIF > 1$ means that the test image has a superior visual quality than the reference image.

3.3.4.6. *Parameter specification in the VIF algorithm*

In order to compute (26), values of $\mathbf{C}_u$, $z_i$, $g_i$, $\sigma_v$, and $\sigma_n$ must be estimated. The estimation of $\mathbf{C}_u$ is done using the wavelet coefficients of the reference image in each subband:

$$\hat{\mathbf{C}}_u = \frac{1}{M_r}\sum_{i=1}^{M_r} c_i c_i^T \qquad (27)$$

By using maximum likelihood estimation, $z_i^2$ can be estimated using the following equation [61]:

$$\hat{z}_i^2 = \frac{1}{M_m} c_i^T \hat{C}_u^{-1} c_i \tag{28}$$

where $M_m$ is the dimensionality of $c$. The parameters $g_i$ and $\sigma_{v,i}$ can be computed using simple regression, since both the reference and test image coefficients are available. Finally, $\sigma_n$ is estimated by running the VIF algorithm for different values of this parameter and then choosing the value that yields the best performance in terms of overall image quality prediction accuracy.

Authors of [10] have provided a MATLAB implementation of the VIF algorithm that is available at [62].

3.3.5. *Most apparent distortion (MAD)*

MAD algorithm [11] assumes that HVS employs different strategies when judging the quality of images. It is mentioned in [11] that when HVS attempts to view images containing near-threshold distortions, it tries to move past the image, looking for distortions. This approach is called detection-based strategy. Moreover, it is also stated in [11] that when HVS attempts to view images containing clearly visible distortions, it tries to move past the distortions, looking for image's subject matter. This approach is called appearance-based strategy. For estimating distortions in detection-based strategy, local luminance and contrast masking are used. Moreover, for estimating distortions in appearance-based strategy, variations in local statistics of spatial frequency components are being employed. Here, we summarize each strategy in more details.

3.3.5.1. *Detection-based strategy*

It is argued in [11] that when HVS views high quality images, it tries to look beyond image's subject matter, looking for distortions. Detection-based strategy consists of two stages: determining the locations of visible distortions, and computing perceived distortion due to visual detection.

First, the locations of visible distortions should be determined. In order to describe the non-linear relationship between pixel values and physical luminance of display device, MAD algorithm primarily transforms pixels of the reference and test images to luminance values using the following equation:

$$L = (\beta + \alpha I)^\gamma \tag{29}$$

where $L$ is the luminance image, $I$ is the reference (or test) image, and $\beta$, $\alpha$, and $\gamma$ are device specific constants. Applying (29) to $I_{ref}$ and $I_{tst}$ yields $L_{ref}$ and $L_{tst}$ respectively. Since HVS has a non-linear response to luminance, it should be converted to perceived luminance via:

$$\hat{L} = L^{\frac{1}{3}} \tag{30}$$

where $\hat{L}$ denotes perceived luminance. Applying (30) to $L_{ref}$ and $L_{tst}$ results in $\hat{L}_{ref}$ and $\hat{L}_{tst}$ respectively. After computing perceived luminance, an error image is computed:

$$\hat{L}_{err} = \hat{L}_{ref} - \hat{L}_{tst}. \tag{31}$$

To describe variations in sensitivity due to spatial frequency, authors of [11] employ contrast sensitivity function (CSF) as introduced in [63] with adjustments as in [64]. CSF is applied to both, the reference and error images which yields $I'_{ref}$ and $I'_{err}$ respectively. Since presence of an image's content can affect the detection of distortions, a spatial domain measure of contrast masking is employed. To model this, first $I'_{ref}$ is divided into blocks of size $16 \times 16$ with 75 percent overlap between neighboring blocks. Afterward, rms contrast (in the lightness domain) of each block is calculated. The rms contrast for block $b$ of $I'_{ref}$ is calculated by:

$$C_{ref}(b) = \frac{\tilde{\sigma}_{ref}(b)}{\mu_{ref}(b)} \tag{32}$$

where $\mu_{ref}(b)$ is the mean of block $b$ in the reference image, and $\tilde{\sigma}_{ref}(b)$ is the minimum of the standard deviation of the four subblocks in $b$. The same procedure is done for $I'_{err}$ with the exception that the rms contrast for this image is calculated using the following equation:

$$C_{err}(b) = \begin{cases} \dfrac{\sigma_{err}(b)}{\mu_{ref}(b)} &, \quad \mu_{ref} > 0.5 \\ 0 &, \quad \text{otherwise} \end{cases} \tag{33}$$

where $\sigma_{err}(b)$ is the standard deviation of block $b$ in $I'_{err}$. In (33), the threshold of 0.5 denotes the fact that HVS is relatively insensitive to changes in extremely dark regions. After computing $C_{ref}(b)$ and $C_{err}(b)$, a local distortion visibility map, $\xi(b)$, is computed as follows:

$$\xi(b) = \begin{cases} \ln(C_{err}(b)) - \ln(C_{ref}(b)) &, \quad \ln(C_{err}(b)) > \ln(C_{ref}(b)) > \delta \\ \ln(C_{err}(b)) - \delta &, \quad \ln(C_{err}(b)) > \delta \geq \ln(C_{ref}(b)) \\ 0 &, \quad \text{otherwise} \end{cases} \tag{34}$$

where $\delta$ is a threshold value ($\delta = -5$, as in [11]).

Second, the perceived distortion due to visual detection ($d_{detect}$) is calculated. $d_{detect}$ is calculated via:

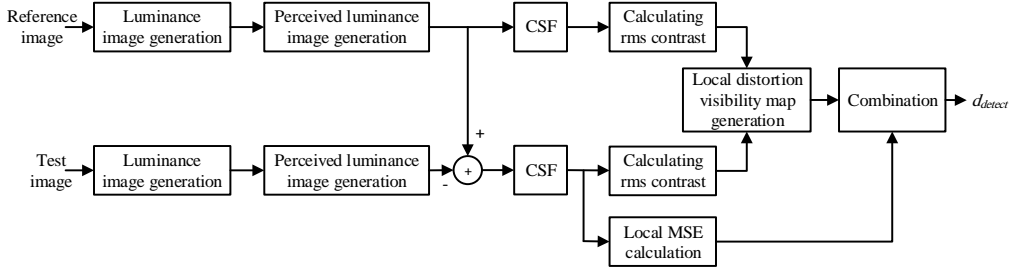

**Fig. 6.** The block diagram of the detection-based strategy in the MAD algorithm.

$$d_{detect} = \left( \frac{1}{B} \sum_b \left[ \xi(b) \times \rho(b) \right]^2 \right)^{\frac{1}{2}} \tag{35}$$

where $B$ is the total number of blocks, and $\rho(b)$ is the local MSE of block $b$ of size $16 \times 16$, that can be calculated using the following equation:

$$\rho(b) = \frac{1}{16 \times 16} \sum_{i,j \in M_p} \left( I'_{err}(i,j) \right)^2 \tag{36}$$

where $M_p$ is the set of pixels in block $b$.

$d_{detect}$ takes its values in the interval $[0, \infty)$. If $d_{detect} = 0$, there are no visible distortions in the test image. As the value of $d_{detect}$ increases, perceived distortion increases and consequently, visual quality decreases. The block diagram of the detection-based strategy is presented in Fig. 6.

3.3.5.2. *Appearance-based strategy*

It is argued in [11] that when viewing low quality images, HVS tries to move past the distortions, looking for image's content. To model this strategy, MAD algorithm uses log-Gabor filter responses. Similar to detection-based strategy, this strategy is also consists of two stages; log-Gabor decomposition of the reference and test images, and computing the local statistical difference map.

First, the reference and test images are decomposed into number of subbands via a 2-D log-Gabor filter bank with frequency responses of the form:

$$G_{s,o}(\omega, \theta) = \exp\left[ -\left[ \frac{\log\left( \frac{\omega}{\omega_s} \right)}{\sqrt{2}\sigma_r} \right]^2 \right] \times \exp\left[ -\frac{(\theta - \theta_0)^2}{2\sigma_0^2} \right] \tag{37}$$

where indices $s$ and $o$ correspond to spatial scale and orientation respectively, parameters $\omega$ and $\theta$ are normalized radial frequency and orientation respectively, $\omega_s$ is normalized center frequency, $\sigma_r$ controls the filter's bandwidth, and $\theta_0$ and $\sigma_0$ are center orientation and angular spread of the filter respectively. In [11], five scales ($s = 1, 2, ..., 5$) and four orientations ($o = 1, 2, ..., 4$) are used for log-Gabor decomposition, which result in 20 subbands per image.

Second, a local statistical difference map, $\eta(b)$, is generated. This map is defined by comparing local subband statistics of the reference image with those of the test image. For each block of size $16 \times 16$, $\eta(b)$ is calculated by:

$$\eta(b) = \sum_{s=1}^{5} \sum_{o=1}^{4} \ell_s \left\{ \left| \sigma_{s,o}^{ref}(b) - \sigma_{s,o}^{tst}(b) \right| \right.$$
$$\left. + 2 \left| \varsigma_{s,o}^{ref}(b) - \varsigma_{s,o}^{tst}(b) \right| + \left| \kappa_{s,o}^{ref}(b) - \kappa_{s,o}^{tst}(b) \right| \right\} \quad (38)$$

where $\sigma_{s,o}(b)$, $\varsigma_{s,o}(b)$, and $\kappa_{s,o}(b)$ correspond to standard deviation, skewness, and kurtosis of $16 \times 16$ subband coefficients associated with scale $s$, orientation $o$, and block $b$. In (38), $\ell_s$ is a scale specific weight which takes into account the preference of HVS for coarser scales over fine ones. (in [11], $\ell_s = 0.5, 0.75, 1, 5,$ and $6$ for finest to coarsest scales, respectively). After computing $\eta(b)$, a final scalar value of perceived distortion, $d_{appear}$, is calculated as follows:

$$d_{appear} = \left( \frac{1}{B} \sum_b \eta^2(b) \right)^{\frac{1}{2}}. \quad (39)$$

$d_{appear}$ takes its values in the interval $[0, \infty)$. If $d_{appear} = 0$, there is no perceived distortion in the test image. As the value of $d_{appear}$ increases, perceived distortion increases and consequently, visual quality decreases. The block diagram of appearance-based strategy is shown in Fig. 7.

After computing $d_{detect}$ and $d_{appear}$, these two values are combined to yield an overall measure of perceived distortion. In [11], it is hypothesized that HVS uses a combination of detection based strategy and appearance based strategy for assessing the quality of images. To model the relation between these two strategies, a weighted geometric mean of $d_{detect}$ and $d_{appear}$ is employed that has the form:

$$\text{MAD} = (d_{detect})^{\alpha} (d_{appear})^{1-\alpha} \quad (40)$$

where $\alpha$ is a weighting constant chosen to reflect the relative importance of each term. MAD measure takes its values in the interval $[0, \infty)$. It is argued that selecting a value for $\alpha$ based on $d_{detect}$ can yield good overall performance [11]. Therefore, $\alpha$ is calculated by:

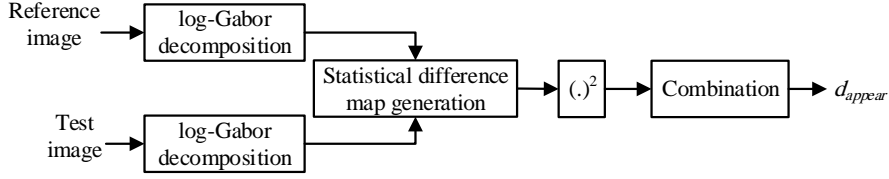

**Fig. 7.** The block diagram of the appearance-based strategy in the MAD algorithm.

$$\alpha = \frac{1}{1+\beta_1 (d_{detect})^{\beta_2}} \qquad (41)$$

where $\beta_1$ and $\beta_2$ are two constants chosen in a way that achieves the best performance of the MAD algorithm in terms of quality prediction accuracy.

3.3.5.3. *Parameter specification in the MAD algorithm*

There are several parameters in the MAD algorithm that need to be specified. First, for computing (29) the values of $\alpha$, $\beta$, and $\gamma$ are set to be 0, 0.02874, and 2.2 respectively. These parameters are calculated using 8 bit pixel values and an sRGB display. Second, the log-Gabor filter parameters are assigned as follows: $\omega_s = \{0.6666, 1.3333, 2, 2.6666, 3.3333\}$ for finer to coarser scales respectively, $\sigma_r = 0.0413$, $\sigma_0 = \pi/6$ rad, and $\theta_0 = \{0, \pi/4, \pi/2, 3\pi/4\}$ rad. Finally, for computing (41) the values of $\beta_1$ and $\beta_2$ are set to be 0.467 and 0.130 respectively. It is important to note that the values of $\beta_1$ and $\beta_2$ are chosen in a way that optimize the performance of the MAD algorithm on the A57 dataset [51].

Authors of [11] have provided a MATLAB implementation of the MAD algorithm that is available at [65].

### 3.3.6. *Feature similarity index (FSIM)*

The FSIM algorithm [12] is based on the fact that HVS understands an image mainly due to its low-level characteristics, e.g., edges and zero crossings [66-68]. In order to assess the quality of an image, FSIM algorithm uses two kinds of features. Physiological and psychophysical experiments have demonstrated that at points with high phase congruency (PC), HVS can extract highly informative features [68-72]. Therefore, PC is used as the primary feature in the FSIM algorithm. However, PC is contrast invariant and our perception of an image's quality is also affected by local contrast of that image. As a result of this dependency, the image gradient magnitude (GM) is used as the secondary feature in the FSIM algorithm. Calculating FSIM measure consists of two stages: computing image's PC and GM, and computing the similarity measure between the reference and test images.

3.3.6.1. *PC and GM computation*

The PC model states that Fourier components with maximum phase contain the points where features are perceived by HVS. This model provides a simple structure on how mammalian visual system handles detection and identification of features in an image [68-72]. First, by applying (37) to the reference and test images, a set of response vectors are created at location $\mathbf{x}$, scale $s$, and orientation $o$. Second, the local amplitude of these vectors at scale $s$ and orientation $o$ is calculated. Moreover, the local energy at orientation $o$ is computed. Finally, the PC value at location $\mathbf{x}$ is calculated using the following equation:

$$PC(\mathbf{x}) = \frac{\sum_o E_o(\mathbf{x})}{\varepsilon + \sum_s \sum_o A_{s,o}(\mathbf{x})} \tag{42}$$

where $E_o(\mathbf{x})$ is the local energy at orientation $o$, $A_{s,o}(\mathbf{x})$ is the local amplitude at scale $s$ and orientation $o$, and $\varepsilon$ is a positive stabilizing constant. $PC(\mathbf{x})$ is a real number that takes its values in the interval $[0,1]$.

In order to compute the gradient magnitude of the reference and test images, three different gradient operators are employed. These operators are: Sobel operator [73], Prewitt operator [73], and Scharr operator [74].

*3.3.6.2. Similarity measure computation*

Consider $PC_{ref}$ and $PC_{tst}$ are $PC$ maps computed for $\mathbf{I}_{ref}$ and $\mathbf{I}_{tst}$ respectively, and $G_{ref}$ and $G_{tst}$ are $GM$ maps for these images. The final similarity measure between the reference and test images consists of two components: similarity measure between $PC_{ref}$ and $PC_{tst}$ or $S_{PC}(\mathbf{x})$, and similarity measure between $G_{ref}$ and $G_{tst}$ or $S_G(\mathbf{x})$. $S_{PC}(\mathbf{x})$ is calculated by the following equation:

$$S_{PC}(\mathbf{x}) = \frac{2 PC_{ref}(\mathbf{x}) PC_{tst}(\mathbf{x}) + T_4}{PC_{ref}^2(\mathbf{x}) + PC_{tst}^2(\mathbf{x}) + T_4} \tag{43}$$

where $T_4$ is a positive stabilizing constant chosen to prevent the denominator from becoming too small. $S_{PC}(\mathbf{x})$ takes its values in the interval $(0,1]$. $S_G(\mathbf{x})$ is calculated by:

$$S_G(\mathbf{x}) = \frac{2 G_{ref}(\mathbf{x}) G_{tst}(\mathbf{x}) + T_5}{G_{ref}^2(\mathbf{x}) + G_{tst}^2(\mathbf{x}) + T_5} \tag{44}$$

where $T_5$ is a positive stabilizing constant. $S_G(\mathbf{x})$ takes its values in the interval $(0,1]$.

The values of $T_4$, and $T_5$ depend on the dynamic range of $PC$ and $GM$ values respectively. The final similarity measure, $S_L(\mathbf{x})$, between $\mathbf{I}_{ref}$ and $\mathbf{I}_{tst}$ is computed as follows:

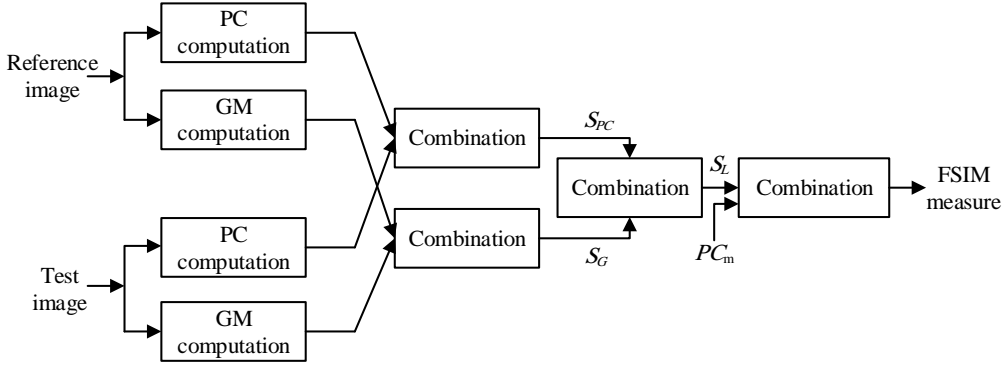

**Fig. 8.** The block diagram of the FSIM algorithm.

$$S_L(\mathbf{x}) = \left[S_{PC}(\mathbf{x})\right]^\alpha \left[S_G(\mathbf{x})\right]^\beta \tag{45}$$

where $\alpha$ and $\beta$ are two constants chosen to indicate the relative importance of each component (in [12], $\alpha = \beta = 1$). Our perception of an image is affected differently by different locations in an image, and a PC value at a location indicates whether that location is perceptually significant or not [72]. Therefore, if either of $PC_{ref}(\mathbf{x})$ and $PC_{tst}(\mathbf{x})$ is greater than the other, it implies that position $\mathbf{x}$ has a higher impact on HVS when evaluating $S_L(\mathbf{x})$ between $\mathbf{I}_{ref}$ and $\mathbf{I}_{tst}$. As a result, FSIM algorithm uses $PC_m(\mathbf{x}) = \max\left(PC_{ref}(\mathbf{x}), PC_{tst}(\mathbf{x})\right)$ as a weighting function for $S_L(\mathbf{x})$ in the overall similarity measure between $\mathbf{I}_{ref}$ and $\mathbf{I}_{tst}$. Finally, the FSIM index between the reference and test images is defined by:

$$\text{FSIM} = \frac{\sum_{\mathbf{x} \in \Omega} S_L(\mathbf{x}) PC_m(\mathbf{x})}{\sum_{\mathbf{x} \in \Omega} PC_m(\mathbf{x})} \tag{46}$$

where $\Omega$ is the whole image spatial domain. The block diagram of the FSIM index is presented in Fig. 8.

3.3.6.3. *Parameter specification in the FSIM algorithm*

In order to specify parameters in the FSIM algorithm, authors of [12] used a subset of the Tampere image dataset 2008 (TID2008) which contained the first 8 reference images and their corresponding 544 test images. Parameters that achieve the highest Spearman's rank order correlation coefficient (SRCC), i.e., a measure of an IQA metric's monotonicity, are chosen, and are fixed for all conducted experiments. In [12], four scales ($s = 1, 2, 3, 4$) and four orientations ($o = 1, 2, 3, 4$) are used for log-Gabor decomposition. The parameters' value in FSIM index are: $\sigma_r = 0.5978$, $\sigma_0 = 0.6545 \text{ rad}$, $T_4 = 0.85$, and $T_5 = 160$. Moreover, $\omega_s = \{1/6, 1/12, 1/24, 1/48\}$ for finer to coarser scales respectively and $\theta_0 = \{0, \pi/4, \pi/2, 3\pi/4\}$ rad. It is mentioned in [12] that Scharr gradient operator [74] yields the highest SRCC among Sobel and Prewitt operators. Therefore, this operator is used to compute GM of the reference and test images.

Authors of [12] have provided a MATLAB implementation of the FSIM algorithm that is available at [75].

## 4  Quality assessment of color images

Objective FR-IQA methods described thus far are designed specifically for gray-scale images and they don't make use of images' color information. Color information simplifies the identification and extraction of objects in a scene. Therefore, it affects human observers' judgment when assessing the quality of an image. In many areas that deal with digital images, there is always a demand for objective quality metrics that can predict the quality of a test color image with respect to its reference version. Applications of such a metric can be found in computer graphics when comparing the level of photorealism of two different rendering methods, image coding when comparing the performance of two different compression schemes, image processing when evaluating the performance of color image enhancement methods, and false-color multispectral image fusion [76]. In general, objective IQA metrics for gray-scale images can, in principle, be extended to incorporate color images. This is accomplished by applying these metrics to each of three RGB color channels individually, and then combining the quality score for each channel together. However, this approach doesn't relate with human perception, and this is because RGB color space doesn't represent color as it is perceived by HVS [76].

The first color image quality measure is proposed in [77]. In this work, a simple model of human color vision is presented which quantitatively describes different perceptual parameters, e.g., brightness and saturation. The perceptual space is considered as a vector space with spatial filtering characteristics. Moreover, a norm on the vector space is introduced that enables measuring the distances and defines a distortion measure that correlates well with perceptual evaluations. Some of the researches that address color image quality assessment can be found in [78-82]. Here, we only describe feature similarity index for color images ($FSIM_C$):

4.1. *Feature similarity index for color images ($FSIM_C$)*

FSIM index described earlier is designed for gray-scale images or the luminance component of color images. In order to extend FSIM index to incorporate color images, first the reference RGB color image is transformed into another color space in which the luminance component can be separated from chrominance. In [12], RGB color image is transformed to YIQ color space, where Y denotes luminance component and I and Q denote chrominance components. RGB color space is transformed to YIQ color space via [83]:

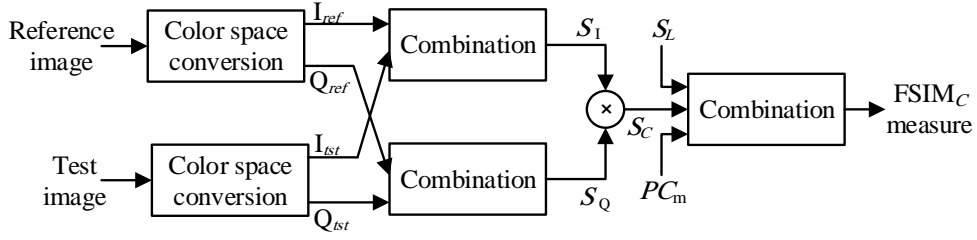

**Fig. 9.** The block diagram of the FSIM$_C$ algorithm.

$$\begin{bmatrix} Y \\ I \\ Q \end{bmatrix} = \begin{bmatrix} 0.299 & 0.587 & 0.114 \\ 0.596 & -0.274 & -0.322 \\ 0.211 & -0.523 & 0.312 \end{bmatrix} \begin{bmatrix} R \\ G \\ B \end{bmatrix}. \quad (47)$$

Suppose $I_{ref}$ and $Q_{ref}$ are chromatic components of the reference image, and $I_{tst}$ and $Q_{tst}$ are chromatic components of the test image. The similarity measures between chromatic components are computed as follows:

$$S_I(\mathbf{x}) = \frac{2I_{ref}(\mathbf{x})I_{tst}(\mathbf{x}) + T_6}{I_{ref}^2(\mathbf{x}) + I_{tst}^2(\mathbf{x}) + T_6} \quad (48)$$

$$S_Q(\mathbf{x}) = \frac{2Q_{ref}(\mathbf{x})Q_{tst}(\mathbf{x}) + T_7}{Q_{ref}^2(\mathbf{x}) + Q_{tst}^2(\mathbf{x}) + T_7} \quad (49)$$

where $T_6$ and $T_7$ are two positive stabilizing constant chosen to prevent the denominators from becoming too small. In [12], the values of $T_6$ and $T_7$ are set to be equal to each other. The final similarity measure between chromatic components, $S_C(\mathbf{x})$, is the product of $S_I(\mathbf{x})$ and $S_Q(\mathbf{x})$:

$$S_C(\mathbf{x}) = S_I(\mathbf{x})S_Q(\mathbf{x}) \quad (50)$$

The FSIM index for color images is calculated by:

$$\text{FSIM}_C = \frac{\sum_{\mathbf{x} \in \Omega} S_L(\mathbf{x})\left[S_C(\mathbf{x})\right]^\lambda PC_m(\mathbf{x})}{\sum_{\mathbf{x} \in \Omega} PC_m(\mathbf{x})} \quad (51)$$

where $\lambda$ is a positive weighting constant chosen to indicate the relative importance of chromatic components. Note that for color images PC and GM are computed by their luminance component Y. Moreover, the calculation process of PC and GM for color images is the same as gray-scale images described in Sec. 3.3.6. The block diagram of the FSIM$_C$ algorithm is presented in Fig. 9.

4.1.1. *Parameter specification in the FSIM$_C$ algorithm*

The values of $s$, $o$, $\omega_s$, $\sigma_r$, $\theta_0$, and $\sigma_0$ in the FSIM$_C$ algorithm are the same as their values in the FSIM algorithm. Moreover, in the FSIM$_C$ algorithm we have: $T_6 = T_7 = 200$, and $\lambda = 0.03$.

Authors of [12] have provided a MATLAB implementation of the FSIM$_C$ algorithm which is available at [75].

## 5 Quality assessment of high dynamic range (HDR) images

There has been a growing interest in recent years in HDR images that have greater dynamic range of intensity values than low dynamic range (LDR) images. In order to visualize HDR images on standard display devices, tone mapping operators (TMOs) [84-87] are employed. Since TMOs reduce the dynamic range of HDR images, they result in information loss and quality degradation. Therefore, it is important to assess the quality of each tone-mapped image to see which TMO provides better quality LDR images. On the other hand, due to the advent of various display technologies, e.g., HDR display, digital cinema projections, and mobile devices' displays, it is important to measure the quality of images with different dynamic ranges to evaluate the capability of each displaying device in producing higher quality images.

Subjective evaluation is the most reliable method for assessing the quality of HDR and LDR images [88-93]. However, as we mentioned before, these methods are expensive, time consuming, and cannot be embedded into optimization algorithms. Therefore, it is important to develop objective IQA methods for evaluating the quality of HDR images and their corresponding tone-mapped versions. The FR-IQA methods described thus far cannot be employed for this purpose. This is due to the fact that the described methods assume that the dynamic range of the reference and test images is similar.

In the following subsections, we will describe two FR-IQA methods for evaluating the quality of images with different dynamic ranges. These methods are: dynamic range independent quality measure (DRIM) [13], designed for evaluating the quality of images with arbitrary dynamic ranges, and tone-mapped images quality index (TMQI) [14], designed for evaluating the quality of tone-mapped images with respect to their reference HDR images.

5.1. *Dynamic range independent quality measure (DRIM)*

In [13], an image quality metric capable of assessing the quality of images with arbitrary dynamic ranges is proposed. The output of this metric is a distortion map that indicates the loss of visible features, the amplification of invisible features, and the reversal of contrast polarity. The DRIM algorithm is sensitive to three types of structural changes:

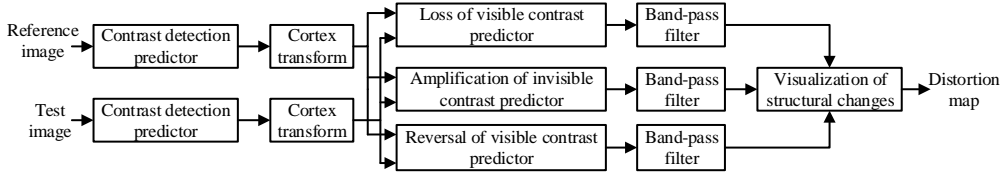

**Fig. 10.** The block diagram of the DRIM algorithm.

- Loss of visible contrast: this case describes the situation in which a contrast that was visible in the reference image becomes invisible in the test image. This usually happens when a TMO compresses the details in the HDR image to a level that they become invisible in the resulting LDR image.
- Amplification of invisible contrast: this case describes the situation where a contrast that was invisible in the reference image becomes visible in the test image. This usually happens when an inverse TMO, i.e., an operator that converts LDR images to HDR images, introduces contouring artifacts in the resulting HDR image.
- Reversal of visible contrast: this case happens when a contrast is visible in both the reference and test images, but with different polarity. This usually occurs in image locations possessing strong distortions.

The block diagram of the DRIM algorithm is presented in Fig. 10. The inputs for this metric are luminance maps corresponding to the reference and test images. First, the detection thresholds are predicted and a perceptually normalized response map is generated. In order to predict detection thresholds, authors of [13] employ the detection model in [94], which is designed specifically for HDR images. This model takes into account spatial sensitivity changes due to local adaption, non-linear response of the photo receptors, and light scattering in the eye's optics. To ensure the accuracy of predictions, the DRIM algorithm calibrates its detection model with measurements in [95]. For optical transfer function (OTF) and CSF, models in [96] and [97] are employed respectively. Second, the perceptually normalized response is decomposed into several bands of different orientations and scales. In order to do this, cortex transform, i.e., the collection of the band-pass and orientation selective filters, as proposed in [97] is employed. Third, for prediction of three distortion types separately for each band, the conditional probability of each distortion type is calculated as follows:

$$P_{loss}^{s,o} = P_{ref/vis}^{s,o} P_{tst/inv}^{s,o} \tag{52}$$

$$P_{ampl}^{s,o} = P_{ref/inv}^{s,o} P_{tst/vis}^{s,o} \tag{53}$$

$$P_{rev}^{s,o} = P_{ref/vis}^{s,o} P_{tst/vis}^{s,o} R^{s,o} \tag{54}$$

where $P_{loss}^{s,o}$, $P_{ampl}^{s,o}$, and $P_{rev}^{s,o}$ denote the conditional probability of loss of visible contrast, amplification of invisible contrast, and reversal of visible contrast in the scale $s$ and orientation $o$ respectively. Subscripts *ref /.* and *tst /.* denote the reference and test images respectively. Also $./vis$ and $./inv$ correspond to visible and invisible contrast. The parameter $R$ is equal to 1 if the contrast polarity in the reference and test images differs, and is zero otherwise. Fourth, because (52) through (54) possess non-linear operators, the probability map $P^{s,o}$ may contain spurious distortions. In order to prevent this problem, each probability map is filtered one more time using its corresponding cortex filter $\boldsymbol{B}^{s,o}$. The filtered probability map is computed as follows:

$$\hat{P}^{s,o}_{loss} = \mathbf{F}^{-1}\left\{\mathbf{F}\left\{P^{s,o}_{loss}\right\}\boldsymbol{B}^{s,o}\right\} \tag{55}$$

where $\mathbf{F}$ and $\mathbf{F}^{-1}$ denote the 2-D Fourier and inverse Fourier transforms respectively. Although (55) is written for $P^{s,o}_{loss}$, the filtered probability maps for $P^{s,o}_{ampl}$ and $P^{s,o}_{rev}$ are computed in a similar manner. Finally, the probability of detecting a distortion in any subband is calculated as follows:

$$P_{loss} = 1 - \prod_{s=1}^{M_s}\prod_{o=1}^{M_o}\left(1-\hat{P}^{s,o}_{loss}\right) \tag{56}$$

where $M_o$ and $M_s$ are total numbers of orientations and scales respectively. Eq. (56) is based on the assumption that detecting each distortion in each subband is an independent procedure. The probability maps $P_{rev}$ and $P_{ampl}$ are calculated in a similar manner.

In order to visualize each of the three distortion types, an in-context distortion map approach similar to [97] is employed, and a custom viewer application for detailed inspections is introduced. In order to generate the in-context map, luminance of the test image is copied to all three RGB channels, and each channel is scaled using the detection probability of their corresponding distortion types. In [13], only the distortion types with highest probability of detection at each pixel location is used for visualization purposes. Green is chosen for loss of visible contrast, blue corresponds to amplification of invisible contrast, and red denotes reversal of visible contrast. By using custom viewer application employed in [13], one can dynamically set the level of distortion types and the background image to an appropriate level in order to investigate each distortion types separately.

Applications of the DRIM algorithm, as stated in [13], are: comparison of TMOs, evaluation of inverse TMOs, and comparison of different types of display devices. To the best of our knowledge, authors of [13] have not published a publicly accessible source code for the DRIM algorithm. However, in [98], authors have provided an online implementation of the DRIM algorithm where the reference and the test images can be uploaded and after assigning the parameters by the user, the probability maps and the final in-context distortion map is generated.

## 5.2. Tone-mapped images quality index (TMQI)

In [14], an objective IQA method for tone-mapped images is proposed. This metric is a combination of multi-scale structural fidelity measure and statistical naturalness measure. The TMQI algorithm consists of two stages: structural fidelity measurement, and statistical naturalness measurement.

Since TMOs compress the dynamic range of HDR images, they result in the loss of information. Moreover, this loss of information may not be visible in the LDR images for the human observers to see. Therefore, structural fidelity is an important part of tone-mapped images quality assessment. Consider $x$ and $y$ to be two local image patches obtained from the HDR and tone-mapped LDR images respectively. TMQI algorithm defines its structural fidelity measure as follows:

$$S_{local}(x,y) = \frac{2\sigma'_x \sigma'_y + T_8}{\sigma'^2_x + \sigma'^2_y + T_8} \frac{\sigma_{x,y} + T_9}{\sigma_x \sigma_y + T_9} \qquad (57)$$

where $\sigma_x$, $\sigma_y$, and $\sigma_{x,y}$ are the local standard deviations and cross correlation between image patches $x$ and $y$ respectively, and $T_8$ and $T_9$ are two positive stabilizing constants designed to prevent the denominators from becoming too small. Compared with (12), the luminance comparison function is missing, and the structure comparison function, denoted by the second part of (57), is exactly the same. The reason for the absence of luminance comparison function is that since TMOs change the local luminance and contrast, the direct comparison of these two characteristics is inappropriate. The first component of (57) is a modified version of (9) that compares the strength of two image signals. This modification is based on two intuitive considerations:

- When the signal strength of the HDR and LDR image patches are either above the visibility threshold or below it, the difference between them should not be penalized.
- The difference in signal strength between HDR and LDR image patches should be penalized when signal strength in one patch is above visibility threshold and is below it in the other patch.

In order to take into account the above considerations, the local standard deviation $\sigma$ is passed through a non-linear mapping that yields $\sigma'$ in (57). This mapping has the following characteristics:

- Signal strengths above visibility threshold are mapped to 1.
- Signal strengths below visibility threshold are mapped to 0.
- Smooth transition between 0 and 1.

The non-linear mapping described above is related to the visual sensitivity of contrast. HVS follows a gradual increasing probability in observing contrast changes. Some psychometric functions that describe the detection probability of signal strength have been used to model the data taken from psychophysical experiments [99,100]. TMQI algorithm employs a commonly used psychometric function known as Galton's ogive [101]. This function has the form of cumulative normal distribution function denoted by:

$$P(a) = \frac{1}{\sigma_a \sqrt{2\pi}} \int_{-\infty}^{a} \exp\left[-\frac{(z-\tau_a)^2}{2\sigma_a^2}\right] dz \tag{58}$$

where $P$ is the probability of detection, $a$ is the amplitude of sinusoidal stimuli, $\tau_a$ is the modulation threshold, and $\sigma_a$ is the standard deviation of normal distribution. It has been shown that $k = \tau_a / \sigma_a$ is approximately a constant, known as Crozier's law [101,102]. Usually, $k$ takes its values between 2.3 and 4, and $k = 3$ results in the probability of detection to be considerably low [101]. In order to quantify visual contrast sensitivity, CSF is being used. TMQI algorithm uses the following equation for CSF [63]:

$$A(f) = 2.6[0.0192 + 0.114f]\exp[-(0.114f)^{1.1}] \tag{59}$$

where $f$ is the spatial frequency. In order for CSF to be compatible with psychological data, it needs to be scaled by a constant $\lambda$. In TMQI algorithm, CSF measurement, as presented in [103], is used. The modulation threshold, $\tau_a(f)$, is calculated via:

$$\tau_a(f) = \frac{1}{\lambda A(f)} \tag{60}$$

Eq. (60) is the threshold value based on contrast sensitivity measurement with assumption of pure sinusoidal stimuli. $\tau_a(f)$ needs to be converted into a signal strength threshold. In order to achieve this, it is important to note that signal amplitude scales with contrast and mean signal intensity. Therefore, the threshold value defined on signal standard deviation, $\tau_s(f)$, is computed as follows:

$$\tau_s(f) = \frac{\mu \tau_a(f)}{\sqrt{2}} = \frac{\mu}{\lambda A(f) \sqrt{2}} \tag{61}$$

where $\mu$ is the mean intensity of the signal. According to Crozier's law [101,102]: $\sigma_s(f) = \tau_s(f)/k$. Finally, the non-linear mapping between $\sigma$ and $\sigma'$ is defined as follows:

$$\sigma' = \frac{1}{\sigma_s \sqrt{2\pi}} \int_{-\infty}^{\sigma} \exp\left[-\frac{(z-\tau_s)^2}{2\sigma_s^2}\right] dz . \tag{62}$$

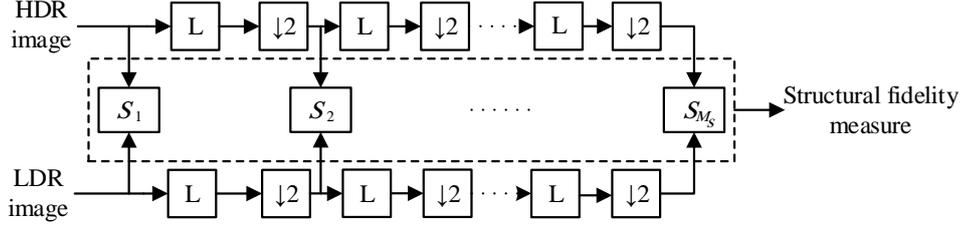

**Fig. 11.** The block diagram of the structural fidelity measure in the TMQI algorithm. L: low-pass filter; ↓2: downsampling by factor of 2.

$\sigma'_x$ and $\sigma'_y$ in (57) are the mapped versions of $\sigma_x$ and $\sigma_y$ respectively. Eq. (57) is computed using a sliding window approach which yields a map containing the variations of structural fidelity across the entire image. TMQI algorithm adapts a multi-scale approach same as MS-SSIM algorithm, in which HDR image and its corresponding LDR version are iteratively low-pass filtered and downsampled (by factor of 2). The block diagram for computing structural fidelity measure is presented in Fig.11. At each scale, the local structural fidelity map is computed and averaged in order to obtain a single score:

$$S_s = \frac{1}{M_p}\sum_{i=1}^{M_p} S_{local}(x_i, y_i) \tag{63}$$

where $x_i$ and $y_i$ are the $i-th$ image patch in the HDR and LDR images respectively, and $M_p$ is the total number of image patches in the scale $s$. The overall structural fidelity score is calculated as follows:

$$S = \prod_{s=1}^{M_s} S_s^{\beta_s} \tag{64}$$

where $M_s$ is the total number of scales, and $\beta_s$ is a constant chosen to indicate the relative importance of the scale $s$.

Structural fidelity alone is not a sufficient measure for evaluating the overall quality of images. Another important characteristic of a high quality LDR image is that it should look natural. According to the results of a subjective experiment conducted in [104], TMQI algorithm uses brightness and contrast for its statistical naturalness model. This model is based on statistics of about 3000 8 bits/pixel gray-scale images available at [105,106]. In order to compute statistical naturalness measure, TMQI algorithm computes the histograms of mean and the standard deviation of these images. It is mentioned in [14] that these histograms can be well-fitted by Gaussian and Beta probability density function respectively:

$$f_m(m) = \frac{1}{\sigma_m\sqrt{2\pi}}\exp\left[-\frac{(m-\bar{m})^2}{2\sigma_m^2}\right] \tag{65}$$

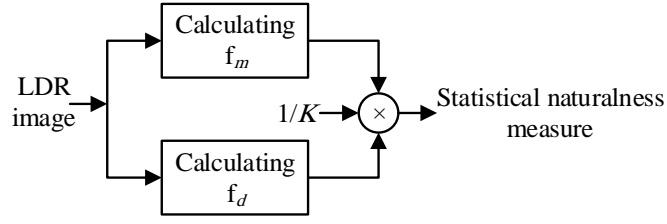

**Fig. 12.** The block diagram of the statistical naturalness measure in the TMQI algorithm.

$$f_d(d) = \frac{(1-d)^{\beta_d-1} d^{\alpha_d-1}}{B(\alpha_d, \beta_d)} \quad (66)$$

where $B(.,.)$ denotes Beta function. According to [107], brightness and contrast are mostly independent characteristics in terms of natural image statistics and biological computation. Therefore, the joint probability density function of contrast and brightness is the product of their respective probability density functions. As a result, TMQI algorithm defines its statistical naturalness measure via following equation:

$$N = \frac{1}{K} f_m f_d \quad (67)$$

where $K = \max\{f_m, f_d\}$ is a normalization factor designed to bound $N$ in the interval $[0,1]$. The block diagram of the statistical naturalness measure is presented in Fig. 12.

After computing structural fidelity and statistical naturalness measure, the overall quality index is calculated via:

$$Q = \gamma S^\alpha + (1-\gamma) N^\beta \quad (68)$$

where $0 \leq \gamma \leq 1$ is a constant chosen to indicate the relative importance of each component, and $\alpha$ and $\beta$ are two constants chosen to indicate each component's sensitivity. The overall quality measure, $Q$, takes its values in the interval $[0,1]$. Two application of the TMQI algorithm, as mentioned in [14], are: parameter tuning in TMOs and adaptive fusion of tone-mapped images.

5.2.1. *Parameter specification in the TMQI algorithm*

There are several parameters in the TMQI algorithm that need to be specified. First, for computing (57) the values of $T_8$ and $T_9$ are set to be 0.01 and 10 respectively. It is mentioned in [14] that the overall performance of the TMQI algorithm is insensitive to the values of parameters $T_8$ and $T_9$ up to an order of magnitude. Second, TMQI algorithm employs the same procedure as SSIM algorithm for creating the fidelity map of each scale, i.e., using a Gaussian sliding window of size $11 \times 11$ and standard deviation of 1.5 samples. Third, the viewing distance is set to be 32 cycles/degree. Therefore, the spatial frequency parameter in (59) is set to be 16 cycles/degree for the finest scale measurement. The spatial frequency parameter employed for the remaining finer scales are 8, 4, 2, and 1 cycles/degree. Fourth, the value of mean intensity in (61) is set to be equal to the dynamic range of LDR images. In other words, $\mu = 128$. Fifth, according to psychophysical experiment in [9], the parameters in (64) are defined as: $M_s = 5$ and $\beta_s = \{0.048,\ 0.2856,\ 0.3001,\ 0.2363,\ 0.1333\}$ for scales 1 to 5 respectively. Finally, since TMQI algorithm is designed specifically for gray-scale images, color images are first converted from RGB color space to Yxy color space and then the structural fidelity measure is applied to the Y component only.

The parameters of (65) and (66) are estimated by first, fitting the histograms of means and standard deviations of images in [105,106] using Gaussian and Beta probability density functions, and then using regression. These parameters, are found to be $\bar{m} = 115.94$ and $\sigma_m = 27.99$ in (65), and $\alpha_d = 4.4$ and $\beta_d = 10.1$ in (66).

Parameters of (68) are determined in a way that best fit the subjective evaluation data presented in [108]. In this subjective experiment, subjects were trained to look simultaneously at two LDR images generated via two different TMOs, and then pick the LDR image that they think has higher overall quality. In order to find the best parameters, an iterative learning method is employed. In this method, at each iteration a pair of images are chosen randomly from a random dataset. If the output of the overall quality measure is of the same order as the subjective rank, then the model parameters are left unchanged. Otherwise, each parameter is updated to lower the difference between subjective and objective scores. The iteration process continues until convergence occurs. It has been reported in [14] that this process has good convergence property. In order to evaluate the robustness of the proposed iterative learning process, a leave-one-out cross validation procedure is employed. It is mentioned in [14] that although this procedure ended up with a different value for parameters at each time, the results were fairly close to one another and they were all of the same rank orders for all datasets. Finally, the parameters of (68), are found to be: $\gamma = 0.8012$, $\alpha = 0.3046$, and $\beta = 0.7088$.

Authors of [14] have provided a MATLAB implementation of the TMQI algorithm that is available at [18].

## 6 Subjective datasets and performance measures in image quality assessment

### 6.1. *Subjective datasets*

In order to evaluate the performance of a newly proposed IQA method, many subjective quality datasets have been introduced. Here, we briefly introduce six most widely used subjective quality datasets. These datasets include: Cornell-A57 dataset [51], IVC dataset [109], Tampere image dataset 2008 (TID2008) [17],

LIVE dataset [15], Toyoma-MICT dataset [110], and categorical image quality (CSIQ) dataset [16].

The Cornell-A57 [51] dataset constitutes of 54 distorted images with six types of distortions. The distortions in this dataset are: quantization of the LH subbands of a 5-level DWT of the images using 9/7 filters, additive white Gaussian noise, baseline JPEG compression, JPEG2000 compression without visual frequency weighting, blurring via a Gaussian filter, and JPEG2000 compression with the dynamic contrast-based quantization algorithm.

The IVC dataset [109] consists of 10 reference images and 185 distorted versions of them. Distortion types in this dataset are: JPEG2000 compression, JPEG compression, blurring, and local adaptive resolution coding.

The TID2008 dataset [17] consists of 1700 test images generated from 25 reference images with 17 distortion types at four different distortion levels. 654 observers from three different countries participated in subjective ratings. Lightening condition, screen size, monitor type, and color gamma are varied between experiments in collecting TID2008 dataset. distortion types in this dataset are: additive Gaussian noise, additive noise in color components is more intensive than its counterpart in the luminance component, masked noise, spatially correlated noise, high frequency noise, impulse noise, Gaussian blur, image denoising, JPEG compression, JPEG2000 compression, transmission errors in JPEG compression, transmission errors in JPEG2000 compression, contrast change, intensity shift, local block-wise distortions of different intensity, and non-eccentricity pattern noise. Quality ratings for each image in TID2008 dataset are reported as mean opinion score (MOS).

The LIVE dataset [15] consists of 29 reference images. Distortions in this dataset are: JPEG compression, JPEG2000 compression, white Gaussian noise, blurring, and fast fading channel distortion of JPEG2000 compressed stream. Total number of distorted images in this dataset is 779. Quality ratings for each image in this dataset are reported as DMOS.

The Toyoma-MICT dataset [110] consists of 14 original images. Totally, it consists of 196 images (168 test images and 28 reference images). Distortions in this dataset are: JPEG and JPEG2000 compression. Method used for subjective rating in this dataset is single stimulus categorical rating. Quality ratings for each image in this dataset are reported as MOS.

The CSIQ dataset [16] consists of 30 reference images each distorted using six types of distortions at four to five distortion levels. Distortions in this dataset are: JPEG and JPEG2000 compression, global contrast decrements, additive white and pink Gaussian noise, and Gaussian blurring. Total number of distorted images in this dataset is 866.

6.2. *Performance measures*

By taking into account the non-linearity of subjective ratings introduced during the subjective experiments, it is necessary to perform a non-linear mapping on the objective scores before measuring the correlation between the subjective and objective scores. According to the video quality experts group (VQEG) research [111], in order to obtain a linear relationship between an objective IQA method's score for an image and its corresponding subjective score, each metric score $x$ is mapped to $q(x)$. The non-linear mapping function $q(x)$ is given by the following equation:

$$q(x) = \beta_1 \left( \frac{1}{2} - \frac{1}{1 + \exp(\beta_2(x - \beta_3))} \right) + \beta_4 x + \beta_5 \tag{69}$$

The parameters $\{\beta_1, \beta_2, \beta_3, \beta_4, \beta_5\}$ are calculated through minimizing the sum of squared differences among the subjective and the mapped scores. In order to compare the performance of a newly proposed IQA method with the existing ones, performance evaluation metrics are used. Here, we describe six commonly used performance measures in IQA:

The Pearson's linear correlation coefficient (PLCC) is the linear correlation coefficient between the predicted MOS (DMOS) and subjective MOS (DMOS). PLCC is a measure of prediction accuracy of an IQA metric, i.e., the capability of the metric to predict the subjective scores with low error. The PLCC can be calculated via following equation:

$$\text{PLCC} = \frac{\sum_{i=1}^{M_d}(q_i - \bar{q})(s_i - \bar{s})}{\left(\sum_{i=1}^{M_d}(q_i - \bar{q})^2\right)^{\frac{1}{2}} \left(\sum_{i=1}^{M_d}(s_i - \bar{s})^2\right)^{\frac{1}{2}}} \tag{70}$$

where $s_i$ and $q_i$ are the subjective score and the mapped score for the $i-th$ image in an image dataset of size $M_d$ respectively, and $\bar{q}$ and $\bar{s}$ are the means of the mapped scores and subjective scores respectively.

The Spearman's rank correlation coefficient (SRCC) is the correlation coefficient between the predicted MOS (DMOS) and the subjective MOS (DMOS). SRCC measures the prediction monotonicity of an IQA metric, i.e., the limit to which the quality scores of a metric agrees with the relative magnitude of the subjective scores. The SRCC can be calculated via following equation:

$$\text{SRCC} = 1 - \frac{6\sum_{i=1}^{M_d} d_i^2}{M_d(M_d^2 - 1)} \tag{71}$$

where $d_i$ is the difference between the $i-th$ image's rank in the objective and subjective experiments. SRCC is independent of any monotonic non-linear mapping between objective and subjective scores.

The Kendall's rank correlation coefficient (KRCC) is a non-parametric rank correlation measure that can be calculated via following equation:

$$\text{KRCC} = \frac{M_c - M_{dc}}{\frac{1}{2} M_d (M_d - 1)} \quad (72)$$

where $M_c$ and $M_{dc}$ are the numbers of concordant and disconcordant pairs in the dataset respectively. Like SRCC, KRCC is a measure of the prediction monotonicity.

The outlier ratio (OR) is defined as the percentage of the number of the predictions outside the interval of ±2 times the standard deviation of the subjective scores. OR can be calculated by the following equation:

$$\text{OR} = \frac{M_d}{M'} \quad (73)$$

where $M'$ is the number of outliers. OR measures the prediction consistency of an IQA metric, i.e., the limit to which the metric maintains the accuracy of its predictions.

The root mean square error (RMSE) can be calculated as follows:

$$\text{RMSE} = \left( \frac{1}{M_d} \sum_{i=1}^{M_d} (q_i - s_i)^2 \right)^{\frac{1}{2}} \quad (74)$$

Like PLCC, RMSE is a measure of prediction accuracy.

The mean absolute error (MAE) can be calculated using the following equation:

$$\text{MAE} = \frac{1}{M_d} \sum_{i=1}^{M_d} |q_i - s_i|. \quad (75)$$

Like PLCC and RMSE, MAE is a measure of prediction accuracy.

A good IQA metric should have higher PLCC, KRCC, and SRCC while having lower RMSE, MAE, and OR.

## 7 Evaluation results

### 7.1. *Evaluation of prediction performance*

In this subsection, we will evaluate the prediction performance of the FR-IQA methods described in previous sections: PSNR, SSIM [8], MS-SSIM [9], VIF [10], MAD [11], FSIM [12], FSIM$_C$ [12], and TMQI [14]. For all these methods, we have used their original MATLAB implementation provided by their respective authors.

**Table 1**

Performance evaluation of 8 FR-IQA algorithms described in this paper.

| | CSIQ dataset | | | | |
|---|---|---|---|---|---|
| | **KRCC** | **SRCC** | **PLCC** | **MAE** | **RMSE** |
| **SSIM** | 0.6907 | 0.8756 | 0.8613 | 0.0991 | 0.1334 |
| **PSNR** | 0.6084 | 0.8058 | 0.8000 | 0.1195 | 0.1575 |
| **MAD** | 0.7970 | 0.9466 | 0.9502 | 0.0636 | 0.0818 |
| **FSIM** | 0.7567 | 0.9242 | 0.9120 | 0.0797 | 0.1077 |
| **VIF** | 0.7537 | 0.9195 | 0.9277 | 0.0743 | 0.0980 |
| **MS-SSIM** | 0.7393 | 0.9133 | 0.8991 | 0.0870 | 0.1149 |

| | LIVE dataset | | | | |
|---|---|---|---|---|---|
| | **KRCC** | **SRCC** | **PLCC** | **MAE** | **RMSE** |
| **SSIM** | 0.7963 | 0.9479 | 0.9449 | 6.9325 | 8.9455 |
| **PSNR** | 0.6865 | 0.8756 | 0.8723 | 10.5093 | 13.3597 |
| **MAD** | 0.8421 | 0.9669 | 0.9675 | 5.2202 | 6.9037 |
| **FSIM** | 0.8337 | 0.9634 | 0.9597 | 5.9468 | 7.6780 |
| **VIF** | 0.8282 | 0.9636 | 0.9604 | 6.1070 | 7.6137 |
| **MS-SSIM** | 0.8045 | 0.9513 | 0.9489 | 6.6978 | 8.6188 |

| | TID2008 dataset | | | | |
|---|---|---|---|---|---|
| | **KRCC** | **SRCC** | **PLCC** | **MAE** | **RMSE** |
| **SSIM** | 0.5768 | 0.7749 | 0.7732 | 0.6547 | 0.8511 |
| **PSNR** | 0.4027 | 0.5531 | 0.5734 | 0.8327 | 1.0994 |
| **MAD** | 0.6445 | 0.8340 | 0.8308 | 0.5562 | 0.7468 |
| **FSIM** | 0.6946 | 0.8805 | 0.8738 | 0.4926 | 0.6525 |
| **VIF** | 0.5860 | 0.7491 | 0.8084 | 0.6000 | 0.7899 |
| **MS-SSIM** | 0.6568 | 0.8542 | 0.8451 | 0.5578 | 0.7173 |

| | CSIQ dataset | | | | |
|---|---|---|---|---|---|
| | **KRCC** | **SRCC** | **PLCC** | **MAE** | **RMSE** |
| **FSIM$_C$** | 0.7690 | 0.9310 | 0.9192 | 0.0762 | 0.1034 |

| | LIVE dataset | | | | |
|---|---|---|---|---|---|
| | **KRCC** | **SRCC** | **PLCC** | **MAE** | **RMSE** |
| **FSIM$_C$** | 0.8363 | 0.9645 | 0.9613 | 5.8403 | 7.5296 |

| | TID2008 dataset | | | | |
|---|---|---|---|---|---|
| | **KRCC** | **SRCC** | **PLCC** | **MAE** | **RMSE** |
| **FSIM$_C$** | 0.6991 | 0.8840 | 0.8762 | 0.4875 | 0.6468 |

| | Dataset in [18] | |
|---|---|---|
| | **KRCC** | **SRCC** |
| **TMQI** | 0.5579 | 0.7385 |

Since the DRIM algorithm [13] doesn't generate a single quality score for the entire image, it is impossible to compare its results with subjective evaluations. Therefore, we have not included this metric in all evaluations. Moreover, since the described FR-IQA methods are for different category of images (some for gray-scale images, some for color images, and some for HDR images), we evaluate the performance of each category separately. The performance evaluation process for TMQI algorithm [14] is done on the dataset presented in [18]. For the remaining algorithms, we choose three datasets, these datasets include: TID2008 dataset [17], LIVE dataset [15], and CSIQ dataset [16]. It is important to note that in all our evaluations, the reference images are excluded and only test images are employed.

Table 1 shows our test results of the 8 FR-IQA methods on four subjective quality datasets. To provide an evaluation of the overall performance of the image quality metrics under consideration, Table 2 gives the average SRCC, KRCC, PLCC, RMSE, and MAE results over three datasets, where the average values are calculated in two cases. In the first case, the performance measures' scores are directly averaged, while in the second case, different weights are assigned to different datasets depending on their sizes (measured as the number of images, i.e., 1700 for TID2008, 866 for CSIQ, and 779 for LIVE datasets respectively). Since TMQI algorithm's performance is measured in only one dataset, it is not included in Table 2.

As it can be seen from Table 1, for TMQI algorithm only SRCC and KRCC measures are calculated. This is due to the fact that PLCC, RMSE, or MAE are used when subjects rank the quality of images in a specific range, e.g., from 1 to 10. However, in the subjective experiment in [18] subjects were asked to rank the images from best to worst quality and thus the scores given by subjects do not represent the quality of images. Hence, only the SRCC and KRCC measures are calculated for evaluation of TMQI algorithm.

7.2. *Evaluation of computation time*

We have also evaluated the computation time of each selected FR-IQA methods. Since authors of [13] have not published a publicly available source code of their algorithm, we have not included DRIM algorithm in our evaluation. As we mentioned before, since the selected methods are for different category of images, we evaluate their computation time separately. We measured the average computation time required to evaluate the quality of images of size $512 \times 512$, Experiments were performed on a laptop with Intel Core i7 processor at 1.6 GHz. The software platform was MATLAB R2013a. The results are listed in Table 3.

**Table 2**

Average performance over three datasets.

| Direct Average | | | | | |
|---|---|---|---|---|---|
| | **KRCC** | **SRCC** | **PLCC** | **MAE** | **RMSE** |
| **SSIM** | 0.6879 | 0.8661 | 0.8598 | 2.5621 | 3.3100 |
| **PSNR** | 0.5659 | 0.7448 | 0.7486 | 3.8205 | 4.8722 |
| **MAD** | 0.7612 | 0.9158 | 0.9162 | 1.9467 | 2.5774 |
| **FSIM** | 0.7617 | 0.9227 | 0.9152 | 2.1730 | 2.8127 |
| **VIF** | 0.7226 | 0.8774 | 0.8988 | 2.2604 | 2.8339 |
| **MS-SSIM** | 0.7335 | 0.9063 | 0.8977 | 2.4475 | 3.1503 |

| Dataset Size-Weighted Average | | | | | |
|---|---|---|---|---|---|
| | **KRCC** | **SRCC** | **PLCC** | **MAE** | **RMSE** |
| **SSIM** | 0.6574 | 0.8413 | 0.8360 | 1.9729 | 2.5504 |
| **PSNR** | 0.5220 | 0.6943 | 0.7017 | 2.9016 | 3.7018 |
| **MAD** | 0.7300 | 0.8941 | 0.8935 | 1.5148 | 2.0085 |
| **FSIM** | 0.7431 | 0.9111 | 0.9037 | 1.6559 | 2.1476 |
| **VIF** | 0.6858 | 0.8432 | 0.8747 | 1.7464 | 2.1999 |
| **MS-SSIM** | 0.7126 | 0.8921 | 0.8833 | 1.8658 | 2.4015 |

| Direct Average | | | | | |
|---|---|---|---|---|---|
| | **KRCC** | **SRCC** | **PLCC** | **MAE** | **RMSE** |
| **FSIM$_C$** | 0.7681 | 0.9265 | 0.9198 | 2.1347 | 2.7599 |

| Dataset Size-Weighted Average | | | | | |
|---|---|---|---|---|---|
| | **KRCC** | **SRCC** | **PLCC** | **MAE** | **RMSE** |
| **FSIM$_C$** | 0.7491 | 0.9149 | 0.9072 | 1.6276 | 2.1090 |

**Table 3**

Evaluation of computation time.

| Computation Time for an image of size $512 \times 512$ (in seconds/image) | | | | | | |
|---|---|---|---|---|---|---|
| | **SSIM** | **PSNR** | **MAD** | **FSIM** | **VIF** | **MS-SSIM** |
| **Time** | 0.0293 | 0.0035 | 2.0630 | 0.3508 | 1.3647 | 0.0834 |

| Computation Time for an image of size $512 \times 512$ (in seconds/image) | |
|---|---|
| | **FSIM$_C$** |
| **Time** | 0.3776 |

| Computation Time for an image of size $512 \times 512$ (in seconds/image) | |
|---|---|
| | **TMQI** |
| **Time** | 0.4087 |

# 8 Quality assessment of 3-D images

The number of digital 3-D images available for human consumption has increased at a fast pace in recent years. According to the statistics collected by the motion picture association of America (MPAA), half of all moviegoers saw at least one 3-D movie in 2011, and those under 25 years old saw more than twice that number [112]. In order to meet this increasing demand, the number of 3-D movies has been increasing at least 50 percent annually over the recent years [112,113]. Aside from movies, other forms of 3-D contents are finding their way into our daily lives via 3-D television broadcasts [114], and 3-D on mobile devices [115]. These contents bring with themselves a variety of complex technological and perceptual problems. For a consistent, comfortable, and plausible perception of depth, a large number of parameters in the imaging and processing stages need to be determined in a perceptually meaningful way. However, due to some inevitable trade-offs in real-world applications, the visual quality of these 3-D contents will degrade. Therefore, in order to maintain and improve the quality of experience (QoE) of 3-D visual contents, subjective and objective quality assessment methods are needed. These methods are of high importance for display manufacturers, content providers, and service providers. Compared to its 2-D counterpart, 3-D IQA faces more new challenges. These include depth perception, virtual view synthesis, and asymmetric stereo compression.

One natural question is the applicability of 2-D IQA methods to the 3-D images. The works in [116,117] try to answer this question. The results demonstrated that 2-D objective IQA methods can well evaluate the quality of 3-D images only in the case of symmetric images, i.e., the PSNR's of the two-eye images are approximately the same.

Some of the proposed quality descriptors of 3-D contents that quantify the overall viewing experience of a 3-D representation are as follows [118]:

- Depth quality: the depth characteristics of 3-D data need to be examined in order to validate the suitability of the content for viewing [119].
- Naturalness: the limit that enables viewers to easily fuse left and right views into a natural-looking 3-D image with smooth depth representation [120].
- Presence: a natural-looking 3-D scene enhances the viewers' sense of presence [121].
- Value-add: the perceived benefit of displaying a content in 3-D over displaying the same content in 2-D [122].
- Discomfort: the overall subjective perception resulting from physiological and/or psychological effects of 3-D viewing content [123].
- Overall 3-D QoE as typically measured in terms of DMOS.

It is important to note that there are no commonly accepted methods for quantifying the above descriptors yet. However, Standards have recently been introduced to address this issue. Here, we summarize some of these standards:

- ITU-R [124] has released a new recommendation on subjective quality assessment of 3-D TV systems. The focus of this recommendation is on picture quality, depth quality, and visual comfort.
- The VQEG is addressing three main areas, including finding ground truth data for subjective evaluation methodology validation, validating objective 3-D video quality evaluation, and determining the effects of viewing environment on 3-D quality assessment.
- IEEE initiated work on a standard for quality assessment of 3-D contents, 3-D displays, and 3-D devices based on human factors. This work looks into characteristics of display, device, environment, content, and viewers.

The classification of 2-D IQA methods (namely FR-IQA, RR-IQA, and NR-IQA) can be used in the case of 3-D images. However, the definitions do not apply in quite the same way [125]. This is due to the fact that it is impossible to gain access to the reference and test 3-D images as they are perceived. This results since we only can access the left and right views of a scene, and we cannot access the Cyclopean image, i.e., a single mental image of a scene generated by the brain through combining the images received from the two eyes. This applies to both, the reference and test Cyclopean images. Therefore, the problem of 3-D IQA is double-blind [125].

The first objective IQA for 3-D images is presented in [126]. In this work, a quality metric which uses the reliable 2-D IQA methods (including SSIM [8], UQI [52,53], method in [45], and the metric in [38]) is proposed. It is noteworthy that this method doesn't take into account the depth information of 3-D images.

Based on utilized information, 3-D IQA methods can be classified in two categories [127]: methods based on color information only, and methods based on color and disparity information.

8.1. *Methods based on color information*

The methods in this category are based solely on color information [128-132]. In [128], quality scores on the SIFT-matched feature points are computed. In [129], a multiple channel model is employed to estimate the 3-D image quality. In [130], an RR-IQA method for 3-D images is proposed. This method makes use of extracted edge information. In [131], the Gabor response of binocular vision is modeled for measuring the quality of 3-D images. In [132], a state of the art 3-D IQA method for 3-D video compression is proposed.

8.2. *Methods based on color and disparity information*

The methods in this category make use of both, color and disparity information in order to evaluate the overall quality of 3-D data [133-135]. In [133], an RR-IQA method for 3-D images is proposed which is based on eigenvalues/eigenvectors analysis. In [134,135], two NR-IQA methods for 3-D images are proposed.

*8.3. Subjective 3-D image quality datasets*

In this subsection, some of the subjective 3-D image quality datasets are introduced:

LIVE 3-D IQA dataset [136] consists of 20 reference image, 5 distortion categories, and total number of 365 test images. The quality scores in this dataset are in the form of DMOS. This dataset is the first publicly available 3-D IQA dataset that makes use of true depth information along with stereoscopic pairs and human opinion scores. Distortion types in this dataset are: JPEG compression, JPEG2000 compression, additive white Gaussian noise, Gaussian blur, and fast fading model based on the Rayleigh fading channel.

IVC 3-D images dataset [137] consists of 6 reference images and 15 distorted version of each image plus their respective subjective scores. The distortion types in this dataset are: JPEG compression, JPEG2000 compression, and blurring. Total number of images in this dataset is 96.

To the best of our knowledge, the only 3-D dataset for HDR images and their corresponding tone-mapped versions is available in [138]. In this dataset, 9 reference 3-D HDR images are tone-mapped using 8 TMOs. The total number of images in this dataset is 81. Moreover, the statistics of these images (including min, max, and mean luminance) and their histograms are also available in this dataset.

## 9 Conclusion

The growing demand for digital image technologies in applications like medical imaging, biomedical systems, monitoring, and communications has highlighted the need for accurate quality assessment methods. Many processes can affect the quality of images, including compression, transmission, display, and acquisition. Therefore, accurate measurement of the image quality is an important step in many image-based applications. IQA aims at quantifying the quality of image signals including 3-D images, retargeted images, and HDR images by means of objective quality metrics. The goal of objective IQA is to design algorithms that can automatically evaluate the quality of images in a perceptually consistent manner. These methods are crucial to multimedia systems since they remove or reduce the need for extensive subjective evaluation.

In this paper, an overview of subjective and objective IQA was presented. Four most commonly used subjective IQA methods were briefly introduced. Moreover, the three main categories of objective IQA were described. 3-D, Color and HDR images quality assessment were also reviewed. The central theme of this study was on FR-IQA methods and we thoroughly described 9 methods of this category. The prediction performance and computation time of these methods were also evaluated.

IQA has been a rapidly developing field of research in recent years. The number of algorithms that are being proposed are growing at a fast pace. Of course, only a small number of methods have been discussed in detail in this paper. The selected methods are widely cited in the literature and have been reported to have good performance by researchers. Another criterion for our selection is that the source code for most of these methods has been made available online, so the interested readers can implement them and regenerate the reported results. There are number of factors that need to be taken into account when selecting an IQA method for a specific application. Some of these factors include the availability of the reference image, computation time, implementation complexity, application goal, and quality prediction accuracy. By considering all these factors, one can make the right choice for each specific application.

We have also provided a brief introduction to 3-D IQA, and summarized the issues associated with this field of research. It is important to note that with the advances in 3-D coding, transmission, and displays, the quality assessment of 3-D images has been studied independently for each of these areas. A number of elements must be taken into account for achieving a 3-D IQA method. Among these are: dependencies between display, content, and the viewer, also individual user constraints, preferences, and perception of depth must be considered. Once we are able to further develop our knowledge of the perception of stereoscopic distortions, we can achieve better 3-D IQA algorithms.